 \documentclass[final,1p,times]{elsarticle}

\pdfoutput=1

\usepackage{amssymb}
\usepackage{amsthm}
\usepackage{amsmath}
\usepackage{multirow}
\usepackage{booktabs}
\usepackage{graphicx}
\usepackage{epstopdf}
\usepackage{pdflscape}
\usepackage{amsthm}
\usepackage{amsmath}
\usepackage[english]{babel}
\usepackage{subfigure}
\usepackage{threeparttable}

\usepackage{graphicx}
\usepackage{subfigure}

\usepackage{tabularx}
\usepackage{graphicx}
\usepackage{threeparttable}
\usepackage[figuresright]{rotating}
\allowdisplaybreaks

\newtheorem{theorem}{Theorem}[section]

\theoremstyle{definition}

\theoremstyle{remark}
\newtheorem{remark}[theorem]{Remark}

\usepackage{xcolor}

\journal{XXXXX}

\begin{document}

\begin{frontmatter}

\title{Very-high-order TENO schemes with adaptive accuracy order and adaptive dissipation control}

\author[a,b,c]{Lin Fu\corref{cor1}}
\ead{linfu@ust.hk}
\cortext[cor1]{Corresponding author.}

\address[a]{Department of Mechanical and Aerospace Engineering, The Hong Kong University of Science and Technology, Clear Water Bay, Kowloon, Hong Kong}
\address[b]{Department of Mathematics, The Hong Kong University of Science and Technology, Clear Water Bay, Kowloon, Hong Kong}
\address[c]{Shenzhen Research Institute, The Hong Kong University of Science and Technology, Shenzhen, China}

\begin{abstract}

In this paper, a new family of very-high-order TENO schemes with adaptive accuracy order and adaptive dissipation control (TENO-AA) is proposed.
The new framework allows for constructing arbitrarily high-order TENO schemes in a unified paradigm and the yielded nonlinear schemes gradually reduce to low-order reconstructions by judging the smoothness with the ENO-like stencil selection strategy. In order to control the nonlinear numerical dissipation adaptively, the flow scales are first measured by examining the first-order undivided difference and the cut-off constant $C_T$ in the TENO weighting strategy is adapted based on the corresponding measurement. With one set of optimal parameters, the newly proposed TENO schemes are designed to deliver excellent performance for predicting highly compressible flows with a wide range of Mach numbers. While the new very-high-order TENO schemes feature good robustness for conventional gas dynamics, the ENO-property is well preserved with the assistant of a positivity-preserving flux limiter for extreme simulations. Without loss of generality, the typical eight- and ten-point TENO-AA schemes are constructed. A set of benchmark simulations are computed to demonstrate the performance of the proposed TENO schemes.

\end{abstract}


\end{frontmatter}


\section{Introduction}

In this paper, we are mainly interested in solving the compressible Euler equations, which belong to the hyperbolic conservation laws, as
\begin{eqnarray}
\label{eq:Euler}
  \frac{{\partial \rho }}{{\partial t}} + \nabla  \cdot (\rho \mathbf{u}) &=& 0 ,\\
  \frac{{\partial (\rho \mathbf{u})}}{{\partial t}} + \nabla  \cdot (\rho \mathbf{u}\mathbf{u} + p\delta ) &=& 0 , \\
  \frac{{\partial E}}{{\partial t}} + \nabla  \cdot (\mathbf{u}(E + p)) &=& 0 ,
\end{eqnarray}
where $\rho$ and $\mathbf{u}$ denote the fluid density and the velocity vector, $E = \rho e + \frac{{\rho \mathbf{u} \cdot \mathbf{u}}}{2}$ is the total energy. To close the equations, the ideal-gas equation of state (EOS) $p = (\gamma  - 1)\rho e$ with $\gamma$ as the ratio of specific heats is employed. The solutions of above governing equations for compressible fluid dynamics typically involve both rich flow length scales and discontinuities, particularly when the flow Mach number is high \cite{shu2009high}\cite{Pirozzoli2011}\cite{fu2021shock}\cite{griffin2021velocity}. In the past decades, many efforts have been dedicated to developing numerical methods, which can resolve the small-scale structures accurately while capturing the discontinuities sharply. Well recognized success includes the developments of the artificial viscosity methods \cite{von1950}\cite{Jameson1994}\cite{Premasuthan2014}, the shock-fitting methods \cite{moretti1966time}\cite{moretti1967three}\cite{moretti1963three}, the total variation diminishing (TVD) methods \cite{van1979towards}\cite{van1977towards}\cite{van1974towards}\cite{Harten1983}\cite{harten1983upstream}, the essentially non-oscillatory (ENO) method \cite{Harten1987} and the weighted essentially non-oscillatory (WENO) methods \cite{Liu1994}\cite{Jiang1996}. In recent years, the high-order WENO schemes become more and more popular, and lots of variants have been proposed to further enhance the performance \cite{shu2016high}\cite{shu2020essentially}.

To improve the accuracy order of the classical WENO-JS scheme \cite{Jiang1996} near the critical points, the fifth-order WENO-M and WENO-Z schemes are developed by Henrick et al. \cite{Henrick2005} and Borges et al. \cite{Borges2008}, respectively. For the purpose of reducing the nonlinear numerical dissipation, Hill and Pullin \cite{Hill2004} propose a switch function to freeze the nonlinear adaptation mechanism when the ratio between the calculated maximum and minimum smoothness indicator is smaller than a problem-dependent threshold. Hu and Adams \cite{Hu2010} propose an adaptive central-upwind sixth-order WENO-CU6 scheme with a modified weighting strategy. Acker et al. \cite{acker2016improved} propose the WENO-Z+ scheme to enhance the performance of WENO schemes by increasing the contribution of the nonsmooth candidate stencils. By minimizing the weighted-integral penalty function in the wavenumber range of interest, the optimized WENO schemes with low numerical dissipation and dispersion errors are proposed in \cite{Weirs1997}\cite{martin2006bandwidth}. Balsara and Shu \cite{balsara2000monotonicity} construct the very-high-order WENO schemes with the monotonicity-preserving technique \cite{Suresh1997} to assure the numerical robustness. Levy et al. \cite{levy1999central}\cite{levy2000compact} propose the central WENO (CWENO) schemes with unequal-sized candidate stencils and a tailored stencil selection procedure. The uniqueness of the CWENO schemes is that the optimal linear weights can be arbitrary positive numbers provided that the sum equals one \cite{capdeville2008central}. More recently, the WENO-AO \cite{balsara2016efficient} and the WENO-ZQ \cite{zhu2016new} scheme are proposed as the cognate variants of CWENO schemes. Other notable contributions in terms of improving the WENO performance with a hybrid concept are made by Adams and Shariff \cite{Adams1996}, Pirozzoli \cite{pirozzoli2002conservative}, and Ren et al. \cite{Ren2003}. More elaborate reviews about the variants of WENO schemes can be found in \cite{shu2016high}\cite{shu2020essentially}.

However, numerical experiments reveal that the performances of classical WENO schemes for fluid simulations involving wide range of spatial and temporal scales are unsatisfactory due to the excessive dissipation produced in the nonsmooth regions \cite{Zhao2014}\cite{fu2019LES}\cite{johnsen2010assessment}\cite{Pirozzoli2011}.
Recently, Fu et al. \cite{fu2016family}\cite{fu2017targeted}\cite{fu2018new}\cite{fu2019LES} propose a family of high-order targeted ENO (TENO) schemes for compressible fluid simulations. The core success of TENO concept is the invention of the ENO-like stencil selection strategy, which either abandons the candidate stencil when crossed by discontinuities or adopts it for the final reconstruction with the optimal linear weight. For smooth flows, the TENO scheme reverts to the optimal background linear scheme exactly instead of asymptotically, distinguishing it from the WENO scheme in a fundamental way. A wide range of benchmark and complex simulations demonstrate that the TENO schemes feature much lower numerical dissipation than the conventional WENO schemes in resolving the small-scale structures while maintaining the ENO property near discontinuities successfully \cite{haimovich2017numerical}\cite{dongdetonation}\cite{fu2019high}\cite{di2020htr}\cite{fu2017targeted}\cite{hamzehloo2021performance}\cite{lusher2020shock}\cite{li2020low}\cite{PENG2021109902}.

In this paper, a new family of very-high-order TENO schemes is proposed attempting to further improve the performance of TENO schemes for multi-scale fluid simulations. The core contributions are as follows: (i) a new framework with increment candidate stencils is first developed. Two types of candidates, i.e. the three three-point stencils of same size and a set of large candidate stencils with incremental width, are involved; (ii) a new targeted ENO-like stencil selection strategy is proposed. For smooth regions, the largest possible candidate which features the best spectral property is applied for the final reconstruction; for non-smooth regions, the reconstruction recursively degenerates to the low-order reconstruction from the high-order reconstruction. The nonlinear adaptation on the three small stencils is invoked only when all large stencils are isolated for the final reconstruction; (iii) based on the measurement of flow scales with the first-order undivided difference, the cut-off parameter $C_T$ is adapted such that small numerical dissipation is produced for resolving the high-wavenumber physical fluctuations while sufficient numerical dissipation for capturing discontinuities; (iv) with one unique set of optimal parameters, benchmark simulations demonstrate that the proposed TENO schemes are low-dissipation with sharp shock-capturing capability for wide range of flow simulations.

The rest of this paper is organized as follows. In section 2, the scalar model equation is introduced briefly; In section 3, the construction of very-high-order TENO schemes with adaptive accuracy order and adaptive dissipation control is elaborated in detail; In section 4, the explicit formulations of the eight- and ten-point TENO-AA schemes are given; In section 5, a set of benchmark simulations including the conventional gas dynamics and the extreme cases are conducted; The performances of the proposed TENO schemes are summarized in the last section.

\section{The scalar model equation}

In the following, the scalar conservation law, i.e.
\begin{equation}
\label{eq:conservationlaw}
\frac{\partial u}{\partial t} + \frac{\partial}{\partial x}f(u) = 0 ,
\end{equation}
where $u$ and $f$ denote the conservative variable and the flux function, respectively, is considered as the model equation of Eq.~(\ref{eq:Euler}). The characteristic velocity is assumed to be $\frac{\partial f(u)}{\partial u} > 0$ for facilitating the presentation. The discretization on a uniform mesh results in a set of ordinary differential equations
\begin{equation}
\label{eq:ordinaryEq}
\frac{d{u_i}}{dt} =  - \frac{\partial f}{\partial x}\left| {_{x = {x_i}}} \right., \text{ } i = 0,\cdots,n.
\end{equation}
The right-hand side of Eq.~(\ref{eq:ordinaryEq}) can be approximated by a conservative finite-difference scheme as
\begin{equation}
\label{eq:implicitEq}
\frac{d{u_i}}{dt} =  - \frac{1}{\Delta x}({h_{i + 1/2}} - {h_{i - 1/2}}) ,
\end{equation}
where the primitive function $h(x)$ is implicitly defined by
\begin{equation}
\label{eq:definitionEq}
f(x) = \frac{1}{\Delta x}\int_{x - \Delta x/2}^{x + \Delta x/2} {h(\xi )d\xi } .
\end{equation}
Eq.~(\ref{eq:implicitEq}) can be further written following
\begin{equation}
\label{eq:approximateEq}
\frac{d{u_i}}{dt} \approx  - \frac{1}{\Delta x}({\widehat f_{i + 1/2}} - {\widehat f_{i - 1/2}}) ,
\end{equation}
where ${\hat f_{i \pm 1/2}}$ denotes the numerical flux.

For standard TENO schemes~\cite{fu2016family}, the numerical flux ${\hat f_{i \pm 1/2}}$ can be computed from a convex combination of $K - 2$ candidate-stencil fluxes as
\begin{equation}
\label{eq:convex}
\widehat f_{i + 1/2} = \sum\limits_{k = 0}^{K - 3} {w_k} \widehat f_{k,i + 1/2}
\end{equation}
for the $K$-point reconstruction. For suppressing Gibbs oscillations \cite{Hewitt1979} in the presence of discontinuities, the weights $w_k$ in Eq.~(\ref{eq:convex}) are computed by the TENO weighting strategy such that the contribution from candidate stencils crossed by discontinuities vanishes in the final reconstruction.

For each candidate, a $(r_k - 1)$-degree interpolation polynomial on the stencil leads to
\begin{equation}
\label{eq:approximatepolynomial}
h(x) \approx {\hat f_k}(x) = \sum\limits_{l = 0}^{r_k - 1} {a_{l,k}} {x^l},
\end{equation}
where $r_k$ denotes the stencil width. After substituting Eq.~(\ref{eq:approximatepolynomial}) into Eq.~(\ref{eq:definitionEq}) and evaluating the integral functions at the stencil nodes, the coefficients ${a_l}$ are determined by solving the resulting system of linear algebraic equations.

\section{The very-high-order TENO-AA scheme}

In this section, we develop a new framework to construct the very-high-order TENO schemes with adaptive accuracy order and adaptive dissipation control in detail. The core ingredients include a new set of candidate stencils, a new targeted ENO-like stencil selection strategy and the adaptive dissipation control strategy.

\subsection{The new framework with incremental candidate stencils}

\begin{figure}[htbp]
\centering
\subfigure{\includegraphics[width=0.9\textwidth]{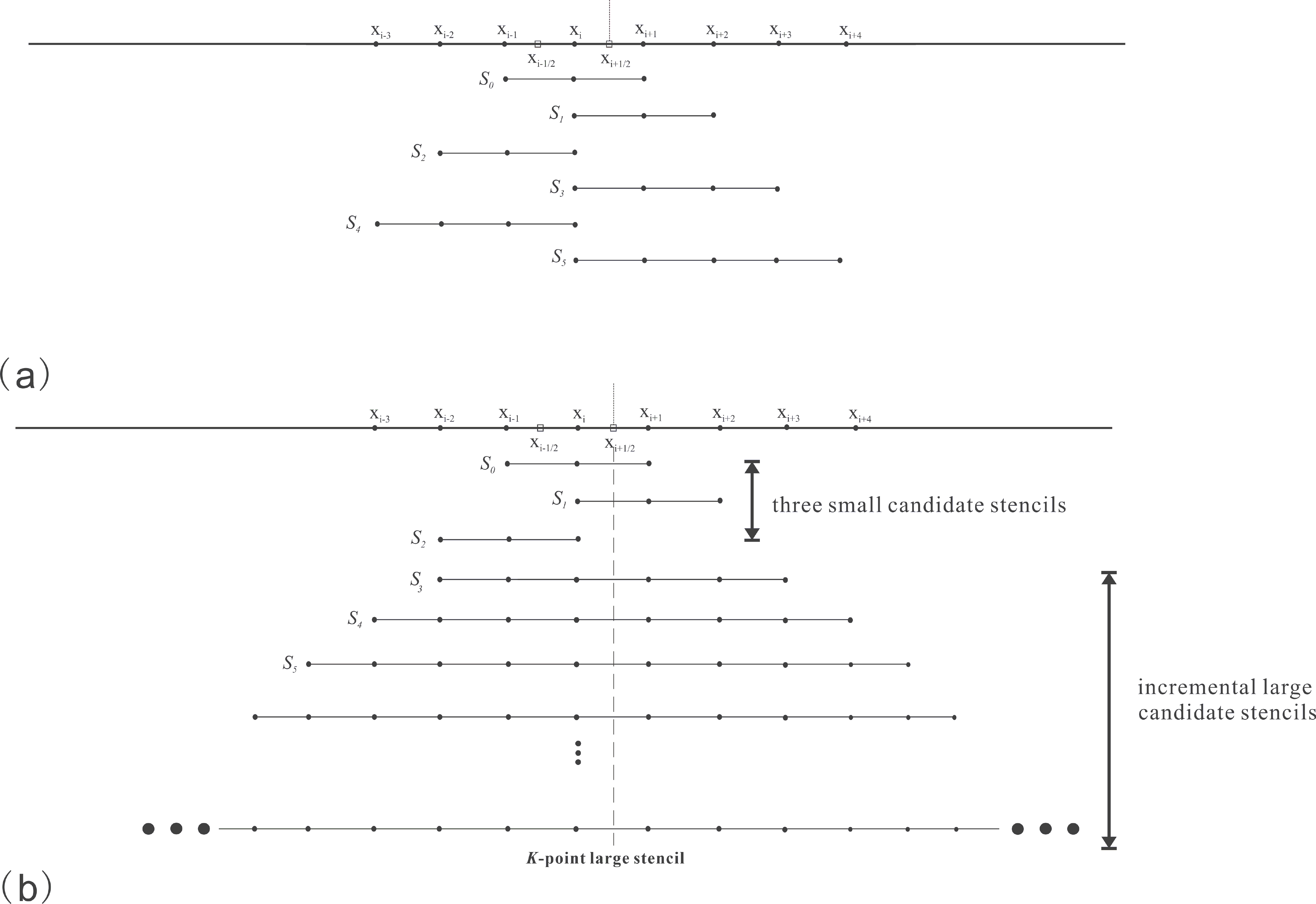}}
\caption{Candidate stencils with incremental width towards high-order reconstruction: (a) the candidate stencils in the standard TENO scheme \cite{fu2016family}; (b) the newly proposed framework for very-high-order TENO reconstruction.}
\label{Fig:incremental_stencil_new}
\end{figure}

As shown in Fig.~\ref{Fig:incremental_stencil_new}, in this paper, we propose a new framework with incremental candidate stencils to construct the very-high-order TENO schemes. There are basically two types of candidate stencils: (i) three three-point candidate stencils of equal size; (ii) symmetric large candidate stencils of incremental width. Based on the experiment that the original five-point TENO5 scheme performs well for compressible gas dynamics \cite{fu2016family}, the first three three-point candidate stencils, which are identical to those of the TENO5 scheme, are designed for robust shock-capturing while the large candidate stencils are incorporated to resolve the smooth flow scales. Considering the large candidate stencils, although the upwind-biased stencils can be involved, they are particularly not recommended since their spectral properties typically do not satisfy the approximate dispersion-relation even with sophisticated optimization methods \cite{Pirozzoli2006}\cite{Hu2014}\cite{Arshed2013}. On the other hand, the dispersion and dissipation errors of symmetric schemes can be optimized flexibly with the relaxation of accuracy order requirement \cite{fu2017targeted}.

\subsection{The strong scale-separation formula}

For the TENO concept, an efficient scale-separation procedure which isolates the discontinuities from smooth flow scales is essentially important for robust shock-capturing \cite{fu2016family}. In this paper, the scale-separation formula is defined as
\begin{equation}
\label{eq:scale_separation}
{\gamma _k}  = \frac{1}{{({\beta _{k,r}} + \varepsilon )}^q}  \text{ , } k = 0, \cdots , p - 1,
\end{equation}
where $\varepsilon = {10^{ - 40}}$ is introduced to avoid the zero denominator and $p$ denotes the total candidate stencil number as in Fig.~\ref{Fig:incremental_stencil_new}. Unlike that in the original WENO-JS schemes \cite{Jiang1996}, $q$ is set as 7 to achieve sufficient scale separation rather than 2. $\beta _{k,r}$, which measures the smoothness of each candidate stencil, can be evaluated following Jiang and Shu \cite{Jiang1996} as
\begin{equation}
\label{eq:measureTENO}
{\beta _{k,r}} = \sum\limits_{j = 1}^{r - 1} {\Delta {x^{2j - 1}}} \int_{{x_{i - 1/2}}}^{{x_{i + 1/2}}} {{{(\frac{{{d^j}}}{{d{x^j}}}{{\hat f}_k}(x))}^2}dx} .
\end{equation}
\subsection{The new targeted ENO-like stencil selection strategy}

For TENO schemes, an ENO-like stencil selection strategy is invented to capture the discontinuity sharply meanwhile maintaining low numerical dissipation in smooth regions \cite{fu2016family}. The basic idea is that one candidate stencil is either abandoned when crossed by discontinuities or applied for the final reconstruction with its optimal weight. In detail, the smoothness indicators are normalized first as
\begin{equation}
\label{eq:normalize}
{{\chi}_k } = \frac{\gamma _k}{{\sum {{\gamma _k}} }} ,
\end{equation}
and then subjected to a sharp cut-off function
\begin{equation}
\label{eq:cutoff}
{\delta _k} = \left\{ {\begin{array}{*{20}{c}}
0, &{\text{if }{\chi}_k  < {C_T},}\\
1, &{\text{otherwise.}}
\end{array}} \right.
\end{equation}
Numerical experiments demonstrate that the cut-off parameter $C_T$ can be optimized to be case-independent \cite{fu2016family}\cite{fu2017targeted}\cite{fu2018new}\cite{fu2019LES}.

In this paper, we develop a new targeted ENO-like stencil selection strategy based on the new framework as shown in Fig.~\ref{Fig:incremental_stencil_new}. With the observation that the larger candidate stencil can feature a better spectral property after optimization, the core idea is to define the largest possible candidate stencil, which is not abandoned by the above ENO-like stencil selection procedure, as the final reconstruction. The flowchart of the reconstruction is elaborated as following
\begin{itemize}

	\item Define a set of candidate stencil groups by assembling the first three small candidates with one large candidate stencil, i.e. ($S_0$, $S_1$, $S_2$, $S_p$), ($S_0$, $S_1$, $S_2$, $S_{p-1}$), $\dots$, and ($S_0$, $S_1$, $S_2$, $S_3$);

    \item Begin with the group with the largest candidate stencil, i.e. ($S_0$, $S_1$, $S_2$, $S_p$), calculate the smoothness indicators of these four candidates with Eq.~(\ref{eq:measureTENO}), perform the scale separation procedure with Eq.~(\ref{eq:scale_separation}), check whether the largest candidate stencil $S_p$ can be applied for the final reconstruction following Eq.~(\ref{eq:normalize}) and Eq.~(\ref{eq:cutoff}). If yes, then return the numerical flux ${{\hat f}_{p,i + 1/2}}$ as the final reconstruction and terminate here; Otherwise, start to consider the left group with the largest possible candidate stencil, i.e. ($S_0$, $S_1$, $S_2$, $S_{p-1}$), $\dots$; so forth and so on;

    \item If no large candidate stencils are chosen for the final reconstruction in above procedure, the first three small candidates ($S_0$, $S_1$, $S_2)$ are considered. The optimal weights $d_k$ subjected to the cut-off $\delta _k$ are re-normalized as
        \begin{equation}
        \label{eq:renormalize}
        {w_k} = \frac{{d_k}{\delta _k}}{{\sum\nolimits_{k = 0}^{2} {d_k}{\delta _k} }} , k = 0, 1, 2
        \end{equation}
        so that the contributions from stencils containing discontinuities vanish eventually. The final reconstructed numerical flux evaluated at cell face $i + \frac{1}{2}$ is assembled
        \begin{equation}
        \label{eq:reconstuction}
        \hat f_{i + 1/2}^K = \sum\limits_{k = 0}^{2} {w_k} {\hat f_{k,i + 1/2}}.
        \end{equation}
\end{itemize}

\begin{remark}
\label{eq:remark_one}
\emph{The first three small candidate stencils play two roles in above reconstruction procedure, i.e. provide the reference to judge whether the large candidate stencil is smooth or not by the ENO-like stencil selection strategy and allow for a certain degree of adaptation for robust shock-capturing.}
\end{remark}

\begin{remark}
\label{eq:remark_two}
\emph{For smooth regions, the largest candidate stencil is assured to be defined for the final reconstruction with the ENO-like stencil selection such that the desired accuracy order can be achieved \cite{fu2016family}. For non-smooth regions, the reconstruction recursively degenerates from high-order to low-order reconstruction until the five-point reconstruction, where the nonlinear adaptation is invoked. In this manner, the new TENO schemes feature the property of adaptive accuracy order.}
\end{remark}

\begin{remark}
\label{eq:remark_two}
\emph{The optimal numerical schemes on the large candidate stencils can be designed by approaching either the maximum accuracy order or the best spectral properties. On the other hand, the optimal weights $d_k, k = 0, 1, 2$ for the small stencils are typically optimized such that the counterpart five-point linear scheme approaches either the highest accuracy order or better spectral property.}
\end{remark}

\subsection{The adaptive dissipation control}

Numerical experiments show that the standard TENO family schemes are still excessively dissipative for resolving the high-wavenumber physical fluctuations \cite{fu2019LES}.
The reason is that the smoothness indicators count for the high-order undivided differences, which are sensitive to both the high-wavenumber physical fluctuations and the genuine discontinuities. Consequently, the small-scale structures are handled by large numerical dissipation since they are not sufficiently isolated from the discontinuities.

Motivated by the work of Ren et al. \cite{Ren2003}, a discontinuity-detecting variable $m$ based on the first-order undivided difference is defined as \cite{fu2019LES}\cite{fu2019very}
\begin{equation}
\label{eq:m_definition}
m = 1 - \min (1,\frac{\eta _{j + 1/2}}{C_r}) ,\\
\end{equation}
where
\begin{equation}
{\eta _j} = \frac{{\left| {2\Delta {f_{j + 1/2}}\Delta {f_{j - 1/2}}} \right| + \varepsilon }}{{{{(\Delta {f_{j + 1/2}})}^2} + {{(\Delta {f_{j - 1/2}})}^2} + \varepsilon }},
\end{equation}
and the parameter $\varepsilon = 10^{-40}$, $C_r = 0.265$ and ${\eta _{j + 1/2}} = \min ({\eta _{j-1}}, {\eta _j},{\eta _{j + 1}},{\eta _{j+2}})$.

To generate adequate numerical dissipation for both discontinuities and high-wavenumber fluctuations, $C_T$ is dynamically adjusted according to the flow scales. The adaptation strategy of $C_T$ is defined as follows
\begin{equation}
\label{eq:CT_definition}
\left\{ {\begin{array}{*{20}{c}}
g(m) = {(1 - m)^4}(1{\rm{ + 4}}m) ,\\
\beta  =  {\alpha _1} - {\alpha _2}(1 - g(m)) ,\\
{C_T} = {10^{-[ \beta ]} } ,
\end{array}} \right.
\end{equation}
where $[\beta]$ denotes the maximum integer which is not larger than $\beta$. In this paper, the parameters are recommended as $\alpha _1 = 14$ and $\alpha _2 = 6.4$. When $m \approx 1$, $C_T$ increases to $10^{-7}$, which is typical for robust shock-capturing with strong nonlinear adaptation. When $m \approx 0$, $C_T \approx 10^{- 14}$, which is suitable for resolving the high-wavenumber physical fluctuations.

Hereafter, the newly proposed TENO family schemes are named as TENO-AA schemes since they feature adaptive accuracy order and adaptive dissipation control.

\section{The eight- and ten-point TENO-AA schemes}

In this section, without loss of generality, we construct the eight- and ten-point TENO-AA schemes in detail. The optimizations of the targeted linear schemes on the candidate stencils are outlined first and the explicit formulas of the TENO8-AA and TENO10-AA schemes are given afterwards.

\subsection{The optimization of the targeted linear schemes on the candidate stencils}

Although an optimal linear scheme with the maximum accuracy order can be straightforwardly constructed on the symmetric stencils by Taylor expansions, the central nature, i.e. with dispersion error whereas without numerical dissipation, renders it inadequate for fluid dynamic simulations as the aliasing dispersion errors can pollute the flow scales for the long-term evolution \cite{Hu2014}. Following Fu et al. \cite{fu2017targeted}, based on the large symmetric candidate stencils in Fig.~\ref{Fig:incremental_stencil_new}, a family of optimal schemes can be constructed with dispersion and dissipation controlled separately as
\begin{equation}
\label{eq:difference_decomposition}
   Df(x) = {D_{dispersion}}f(x) + {D_{dissipation}}f(x) , \\ \hfill
\end{equation}
where
\begin{equation}
\label{eq:dispersion}
{D_{dispersion}}f(x) = \frac{1}{{\Delta x}}\sum\limits_{j = 1}^m {{a_j}(f(x + j\Delta x) - f(x - j\Delta x))}
\end{equation}
without dissipation errors and
\begin{equation}
\label{eq:dissipation}
{D_{dissipation}}f(x) = \frac{1}{{\Delta x}}({b_0}f(x) + \sum\limits_{j = 1}^m {{b_j}(f(x + j\Delta x) + f(x - j\Delta x))} )
\end{equation}
without phase errors. The basic idea is to minimize the dispersion error in terms of certain penalty function while generating adequate numerical dissipation with the payment of accuracy order.

The coefficients $a_j$ in ${D_{dispersion}}f(x)$ can be determined by enforcing the accuracy order constraint and minimizing the weighted-error integral function as
\begin{equation}
\label{eq:weighted-error_integral}
E = \int\limits_0^\pi  {e^{\nu (\pi  - \xi )}} {({\tilde \xi _R} - \xi )^2}d\xi ,
\end{equation}
where the parameter $\nu$ determines the relative importance of error between the low-wavenumber and the high-wavenumber range and ${\tilde \xi _R}$ denotes the real part of the modified wavenumber.

On the other hand, the coefficients $b_j$ in ${D_{dissipation}}f(x)$ can be determined by enforcing the accuracy order constraint and the approximate dispersion relation as \cite{Hu2014}
\begin{equation}
\centering
\zeta = \frac{\left| {\frac{d{\tilde \xi _R}}{d \xi } - 1} \right|}{ - {\tilde \xi _I}} \approx {\cal O}(10) ,
\label{eq:ADR}
\end{equation}
which estimates the necessary dissipation to damp spurious high-wavenumber waves and ${\tilde \xi _I}$ denotes the imaginary part of the modified wavenumber.

For the ten-point scheme on the stencil $S_5$, by constraining the eighth-order accuracy, the optimal reconstructed scheme at cell interface is
\begin{eqnarray}
\centering
\label{eq:10-point}
{{\hat f}_{5,i + 1/2}} & = & 0.001911786299492748{f_{i - 4}} - 0.0170332977472532{f_{i - 3}} \nonumber \\
& + & 0.07339445614860979{f_{i - 2}} -  0.221228429084247{f_{i - 1}} \nonumber \\
& + & 0.6657332621611695{f_i} + 0.6557332621611695{f_{i + 1}}  \nonumber \\
& - &  0.2145617624175796{f_{i + 2}} + 0.07053731329146805{f_{i + 3}} \nonumber \\
& - &  0.01631901203296621{f_{i + 4}} + 0.001832421220128962{f_{i + 5}}.
\end{eqnarray}
Similarly, for the eight-point scheme on the stencil $S_4$, by constraining the sixth-order accuracy, the optimal reconstructed scheme is given as
\begin{eqnarray}
\centering
\label{eq:8-point}
{{\hat f}_{4,i + 1/2}} & = & - 0.006866688980568011{f_{i - 3}} + 0.05128582585522106{f_{i - 2}} \nonumber \\
& - & 0.1968478198727312{f_{i - 1}} + 0.6552858258552222{f_i}  \nonumber \\
& + & 0.6452858258552218{f_{i + 1}} - 0.1908478198727314{f_{i + 2}} \nonumber \\
& + & 0.04928582585522098{f_{i + 3}} - 0.006580974694853729{f_{i + 4}} , \nonumber \\
\end{eqnarray}
and for the six-point scheme on the stencil $S_3$, by constraining the fourth-order accuracy, the optimal reconstructed scheme is given as
\begin{eqnarray}
\centering
\label{eq:6-point}
{{\hat f}_{3,i + 1/2}} & = &  0.02852274270130377 {f_{i - 2}} - 0.1714015614372447 {f_{i - 1}} \nonumber \\
& + & 0.650378818735941 {f_i}  + 0.6253788187359414 {f_{i + 1}}  \nonumber \\
& - & 0.1589015614372448 {f_{i + 2}} + 0.02602274270130375 {f_{i + 3}} . \nonumber \\
\end{eqnarray}

Concerning the optimization of the counterpart linear scheme for the candidate stencil group ($S_0$, $S_1$, $S_2$), the readers are referred to \cite{Arshed2013} for details. The optimal weights are adopted as $d_0 = 0.5065006634, d_1 = 0.3699651429, d_2 = 0.1235341937 $.

%
\subsection{The ten-point TENO-AA scheme}

For simplicity, the derivation of smoothness indicators is based on the assumption that the interpolation polynomials on the candidate stencils follow the formula of Eq.~\ref{eq:approximatepolynomial}. For the ten-point reconstruction, the corresponding smoothness indicators are
\begin{eqnarray}
\label{equ:TENO10_AA}
{\beta _{0,3}} & = & \frac{1}{4}{({f_{i - 1}} - {f_{i + 1}})^2} + \frac{{13}}{{12}}{({f_{i - 1}} - 2{f_i} + {f_{i + 1}})^2} , \nonumber \\
{\beta _{1,3}} & = & \frac{1}{4}{(3{f_i} - 4{f_{i + 1}} + {f_{i + 2}})^2} + \frac{{13}}{{12}}{({f_i} - 2{f_{i + 1}} + {f_{i + 2}})^2} , \nonumber \\
{\beta _{2,3}} & = & \frac{1}{4}{({f_{i - 2}} - 4{f_{i - 1}} + 3{f_i})^2} + \frac{{13}}{{12}}{({f_{i - 2}} - 2{f_{i - 1}} + {f_i})^2} , \nonumber \\
{\beta _{3,6}} & = & \frac{1}{120960}\left[271779 f^2_{i-2}\right. + f_{i-2}(-2380800f_{i-1}
+ 4086352f_{i} - 3462252f_{i+1} \nonumber \\
& + & 1458762f_{i+2} - 245620f_{i+3})
 + f_{i-1}(5653317 f_{i-1} - 20427884 f_{i} \nonumber \\
& + & 17905032f_{i+1} - 7727988f_{i+2} + 1325006f_{i+3}) \nonumber \\
& + & f_{i}(19510972f_{i} - 35817664f_{i+1} + 15929912f_{i+2} - 2792660f_{i+3}) \nonumber \\
& + & f_{i+1}(17195652f_{i+1} - 15880404f_{i+2} + 2863984f_{i+3}) \nonumber \\
& + & \left. f_{i+2}(3824847f_{i+2} - 1429976f_{i+3}) + 139633f^2_{i+3}\right] . \nonumber \\
{\beta _{4,8}} & = & \frac{1}{62270208000}[{f_{i + 4}}(75349098471{f_{i + 4}} - 1078504915264{f_{i + 3}} + 3263178215782{f_{i + 2}} \nonumber \\
  & - & 5401061230160{f_{i + 1}} + 5274436892970{f_i}- 3038037798592{f_{i - 1}}  \nonumber \\
  & + & 956371298594{f_{i - 2}} -  127080660272{f_{i - 3}}) + {f_{i + 3}}(3944861897609{f_{i + 3}}  \nonumber \\
  & - & 24347015748304{f_{i + 2}} +  41008808432890{f_{i + 1}} - 40666174667520{f_i}  \nonumber \\
  & + & 23740865961334{f_{i - 1}} - 7563868580208{f_{i - 2}}  +  1016165721854{f_{i - 3}}) \nonumber \\
  & + & {f_{i + 2}}(38342902371231{f_{i + 2}} - 131738583368480{f_{i + 1}}  +  133017910915250{f_i} \nonumber \\
  & - & 78915800051952{f_{i - 1}} + 25505661974314{f_{i - 2}} - 3471156679072{f_{i - 3}}) \nonumber \\
  & + & {f_{i + 1}}(115593819531025{f_{i + 1}}  - 238366288500400{f_i} + 144163939468910{f_{i - 1}}   \nonumber \\
  & - & 47407534412640{f_{i - 2}} + 6553080547830{f_{i - 3}}) + {f_i}(125743620342175{f_i}   \nonumber \\
  & - & 155660468395520{f_{i - 1}} + 52279668813670{f_{i - 2}} - 7366325742800{f_{i - 3}}) \nonumber \\
  & + & {f_{i - 1}}(49429163447121{f_{i - 1}} - 34065661645264{f_{i - 2}}  + 4916835566842{f_{i - 3}}) \nonumber \\
  & + & {f_{i - 2}}(6047605530599{f_{i - 2}} - 1799848509664{f_{i - 3}}) + 139164877641{f_{i - 3}}{f_{i - 3}}]  . \nonumber \\
{\beta _{5,10}} & = & \frac{1}{8002967132160000}[ \nonumber \\
  &   &{f_{i + 5}}(10271226852556519{f_{i + 5}} - 188781685190428812{f_{i + 4}} + 763744975064517708{f_{i + 3}} \nonumber \\
  & - & 1783557622547881452{f_{i + 2}} + 2647293374939748168{f_{i + 1}} - 2588349166844242788{f_i} \nonumber \\
  & + & 1666621903133989812{f_{i - 1}} - 681536822065683108{f_{i - 2}} + 160665441665569002{f_{i - 3}}  \nonumber \\
  & - & 16642851860701568{f_{i - 4}}) + {f_{i + 4}}(878723894536476309{f_{i + 4}} \nonumber \\
  & - & 7196903757161570412{f_{i + 3}} + 16998197062674090348{f_{i + 2}} \nonumber \\
  & - & 25496314580012763492{f_{i + 1}} + 25171844359828276512{f_i} \nonumber \\
  & - & 16354640170583096868{f_{i - 1}} +  6744593557304053812{f_{i - 2}} \nonumber \\
  & - & 1602760184146136448{f_{i - 3}} + 167317608214622742{f_{i - 4}}) \nonumber \\
  & + & {f_{i + 3}}(14917582508640112584{f_{i + 3}} - 71318426816839529952{f_{i + 2}} \nonumber \\
  & + & 108212250511882175688{f_{i + 1}} - 108001260086756280408{f_i} \nonumber \\
  & + & 70886485695552144912{f_{i - 1}} - 29512192424763178368{f_{i - 2}} \nonumber \\
  & + & 7076170986963512652{f_{i - 3}} - 745034101222016988{f_{i - 4}}) \nonumber \\
  & + & {f_{i + 2}}(86303482562247542424{f_{i + 2}} - 265161231242943624792{f_{i + 1}} \nonumber \\
  & + & 267840170248801834152{f_i} - 177812140823999192448{f_{i - 1}} \nonumber \\
  & + & 74826544942103527152{f_{i - 2}} - 18122888431675718748{f_{i - 3}} \nonumber \\
  & + & 1926367559931410892{f_{i - 4}}) + {f_{i + 1}}(206327958083635665564{f_{i + 1}}  \nonumber \\
  & - & 422297341631652106368{f_i} + 283934849707927541592{f_{i - 1}} \nonumber \\
  & - & 120940245081713867688{f_{i - 2}} + 29628242804292093792{f_{i - 3}}   \nonumber \\
  & - & 3183420029990528028{f_{i - 4}}) + {f_i}(219076505295455164644{f_i}  \nonumber \\
  & - & 298732736184126553512{f_{i - 1}} + 128985549388044842808{f_{i - 2}} \nonumber \\
  & - & 32013238878488743452{f_{i - 3}} + 3482351360282643768{f_{i - 4}}) \nonumber \\
  & + & {f_{i - 1}}(103357453544031038184{f_{i - 1}} - 90608758883967297312{f_{i - 2}} \nonumber \\
  & + & 22823791389362998548{f_{i - 3}} - 2518379721362611092{f_{i - 4}}) \nonumber \\
  & + & {f_{i - 2}}(20182234533315641784{f_{i - 2}} - 10338186802053144852{f_{i - 3}} \nonumber \\
  & + & 1159763060479463988{f_{i - 4}}) + {f_{i - 3}}(1348293007163766699{f_{i - 3}} \nonumber \\
  & - & 308382340247963892{f_{i - 4}}) + 18029727887840089{f_{i - 4}}{f_{i - 4}}] .
\end{eqnarray}

The explicit candidate fluxes are given as
\begin{eqnarray}
\label{equ:TENO10_888stencil}
{{\hat f}_{0,i + 1/2}} & = & \frac{1}{6}( - {f_{i - 1}} + 5{f_i} + 2{f_{i + 1}}) , \nonumber \\
{{\hat f}_{1,i + 1/2}} & = & \frac{1}{6}(2{f_i} + 5{f_{i + 1}} - {f_{i + 2}}) , \nonumber \\
{{\hat f}_{2,i + 1/2}} & = & \frac{1}{6}(2{f_{i - 2}} - 7{f_{i - 1}} + 11{f_i}) , \nonumber \\
{{\hat f}_{3,i + 1/2}} & = &  0.02852274270130377 {f_{i - 2}} - 0.1714015614372447 {f_{i - 1}} \nonumber \\
& + & 0.650378818735941 {f_i}  + 0.6253788187359414 {f_{i + 1}},  \nonumber \\
& - & 0.1589015614372448 {f_{i + 2}} + 0.02602274270130375 {f_{i + 3}} , \nonumber \\
{{\hat f}_{4,i + 1/2}} & = & - 0.006866688980568011{f_{i - 3}} + 0.05128582585522106{f_{i - 2}} \nonumber \\
& - & 0.1968478198727312{f_{i - 1}} + 0.6552858258552222{f_i}  \nonumber \\
& + & 0.6452858258552218{f_{i + 1}} - 0.1908478198727314{f_{i + 2}} \nonumber \\
& + & 0.04928582585522098{f_{i + 3}} - 0.006580974694853729{f_{i + 4}} , \nonumber \\
{{\hat f}_{5,i + 1/2}} & = & 0.001911786299492748{f_{i - 4}} - 0.0170332977472532{f_{i - 3}} \nonumber \\
& + & 0.07339445614860979{f_{i - 2}} -  0.221228429084247{f_{i - 1}} \nonumber \\
& + & 0.6657332621611695{f_i} + 0.6557332621611695{f_{i + 1}}  \nonumber \\
& - &  0.2145617624175796{f_{i + 2}} + 0.07053731329146805{f_{i + 3}} \nonumber \\
& - &  0.01631901203296621{f_{i + 4}} + 0.001832421220128962{f_{i + 5}}.
\end{eqnarray}
\subsection{The eight-point TENO-AA scheme}

For the eight-point TENO8-AA scheme, due to the recursive nature of proposed framework, the reconstruction procedure is identical to that of the ten-point TENO10-AA scheme except that the ten-point candidate stencil is not involved.

\subsection{Implementation discussions}

{The present TENO schemes with wide stencils feature two prominent properties: (1) the arithmetic operation is more intensive; (2) the internode memory communication is larger for distributed-memory computing system when compared to other low-order schemes with  narrower stencils. Considering the fact that the modern high-performance computing system is mainly constructed based on GPU rather than CPU, these two properties would not be the parallel performance bottleneck since the hardware design of GPU with many threads is beneficial for handling the arithmetic intensive operations. In Table 1 of \cite{lusher2021assessment}, it is shown that the computational efficiency of TENO schemes can be significantly improved when GPU is adopted instead of CPU. Moreover, because a single GPU is more powerful than CPU, the necessary partition number to handle a similar scale of computation can be greatly reduced. Therefore, the total internode memory communication is also rather limited, even with the present TENO schemes. The large-scale parallel efficiency of high-order TENO schemes has been demonstrated in \cite{di2020htr}, where the code scales satisfactory up to 128 nodes (i.e., 512 GPUs) with an efficiency of $96.6\%$ on Lassen supercomputer in the weak scaling tests.}

\section{Validations}

In this section, the linear advection and Euler equations are considered. The Roe average is applied for the characteristic decomposition of the convection terms at the cell interface to avoid the spurious oscillations. The Rusanov scheme \cite{Rusanov1961} is employed as the flux splitting method for the conventional compressible gas dynamics while the local Lax-Friedrichs (LLF) \cite{Jiang1996} scheme is adopted for extreme simulations by default. The third-order strong stability-preserving (SSP) Runge-Kutta method \cite{gottlieb2001strong}
\begin{equation}
\label{eq:Runge-kutta}
\begin{array}{l}
{u^{(1)}} = {u^n} + \Delta tL({u^n}) , \\
{u^{(2)}} = \frac{3}{4}{u^n} + \frac{1}{4}{u^{(1)}} + \frac{1}{4}\Delta tL({u^{(1)}}) , \\
{u^{n + 1}} = \frac{1}{3}{u^n} + \frac{2}{3}{u^{(2)}} + \frac{2}{3}\Delta tL({u^{(2)}})
\end{array}
\end{equation}
is deployed for the time integration with a typical CFL number of 0.4, if not mentioned otherwise. For all simulations, the built-in parameters of both proposed schemes are not tuned case-by-case. For the two- and three-dimensional problems, the TENO reconstruction is performed in a dimension-by-dimension manner based on the standard finite-difference framework.

\subsection{Linear advection problem}

The one-dimensional Gaussian pulse advection problem \cite{yamaleev2009systematic} is considered to verify the accuracy order of proposed schemes. The linear advection equation
\begin{equation}
\label{eq:linear_advect}
\frac{\partial u}{\partial t} + \frac{\partial u}{\partial x} = 0 ,
\end{equation}
with initial condition
\begin{equation}
\label{eq:linear_advect_init}
{u_0}(x) = {e^{ - 300(x - {x_c})^2}}, \text{ }  x_c = 0.5 ,
\end{equation}
is solved in a computational domain $ 0 \leq x \leq 1$. Periodic boundary condition is imposed at $ x = 0$ and $x = 1$. The advancing time step is sufficiently small such that the time integration error does not dominate the spatial discretization error. The solution is advanced to $t = 1$ corresponding to one period in time and a sequence of globally refined uniform grids is employed to investigate the convergence in terms of the $L_\infty$ norm.

\begin{figure}%
\centering
  \includegraphics[width=0.98\textwidth]{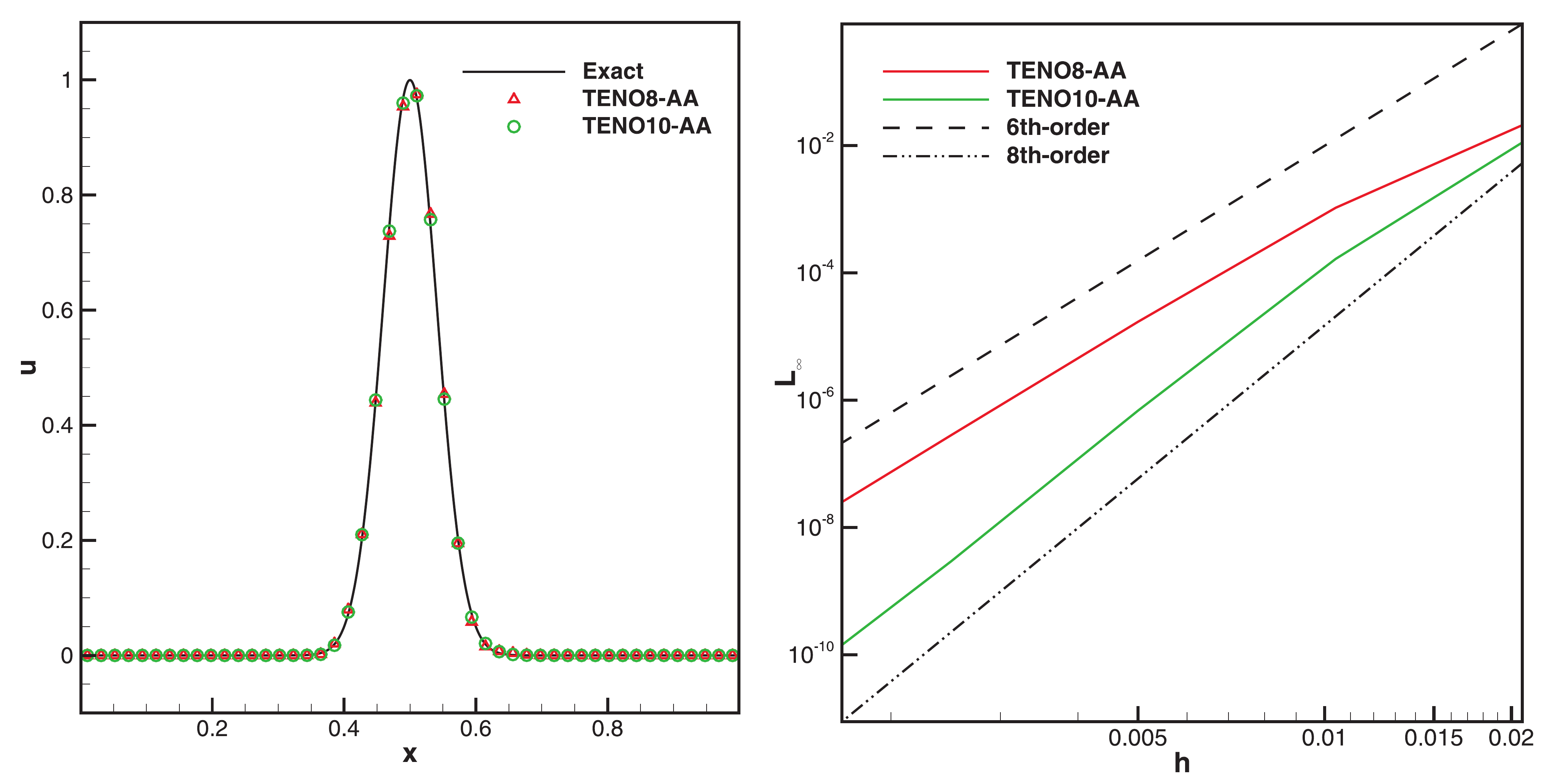}
\caption{The resolved profiles with 48 mesh cells (left) and convergence of the $L_\infty$ error (right) from the TENO8-AA and TENO10-AA schemes.}
\label{Fig:norm_error}
\end{figure}

As shown in Fig.~\ref{Fig:norm_error}, the computed solutions from TENO8-AA and TENO10-AA agree well with the reference and the desired accuracy order is achieved for both schemes.
\subsection{Conventional gas dynamics}
\subsection{Shock tube problem}

The initial condition for the Sod's problem \cite{Sod1978} is
\begin{equation}
\label{eq:sod}
(\rho ,u,p) = \left\{ {\begin{array}{*{20}{c}}
{(1,0,1),}&{\text{if }0 \le x < 0.5 ,} \\
{(0.125,0,0.1),}&{\text{if }0.5 \le x \le 1 ,}
\end{array}} \right.
\end{equation}
and the final simulation time is $t = 0.2$.

The initial condition for the Lax's problem \cite{Lax1954} is
\begin{equation}
\label{eq:lax}
(\rho ,u,p) = \left\{ {\begin{array}{*{20}{c}}
{(0.445,0.698,3.528),}&{\text{if }0 \le x < 0.5 ,} \\
{(0.5,0,0.5710),}&{\text{if }0.5 \le x \le 1 ,}
\end{array}} \right.
\end{equation}
and the final simulation time is $t = 0.14$.

\begin{figure}%
\centering
  \includegraphics[width=0.9\textwidth]{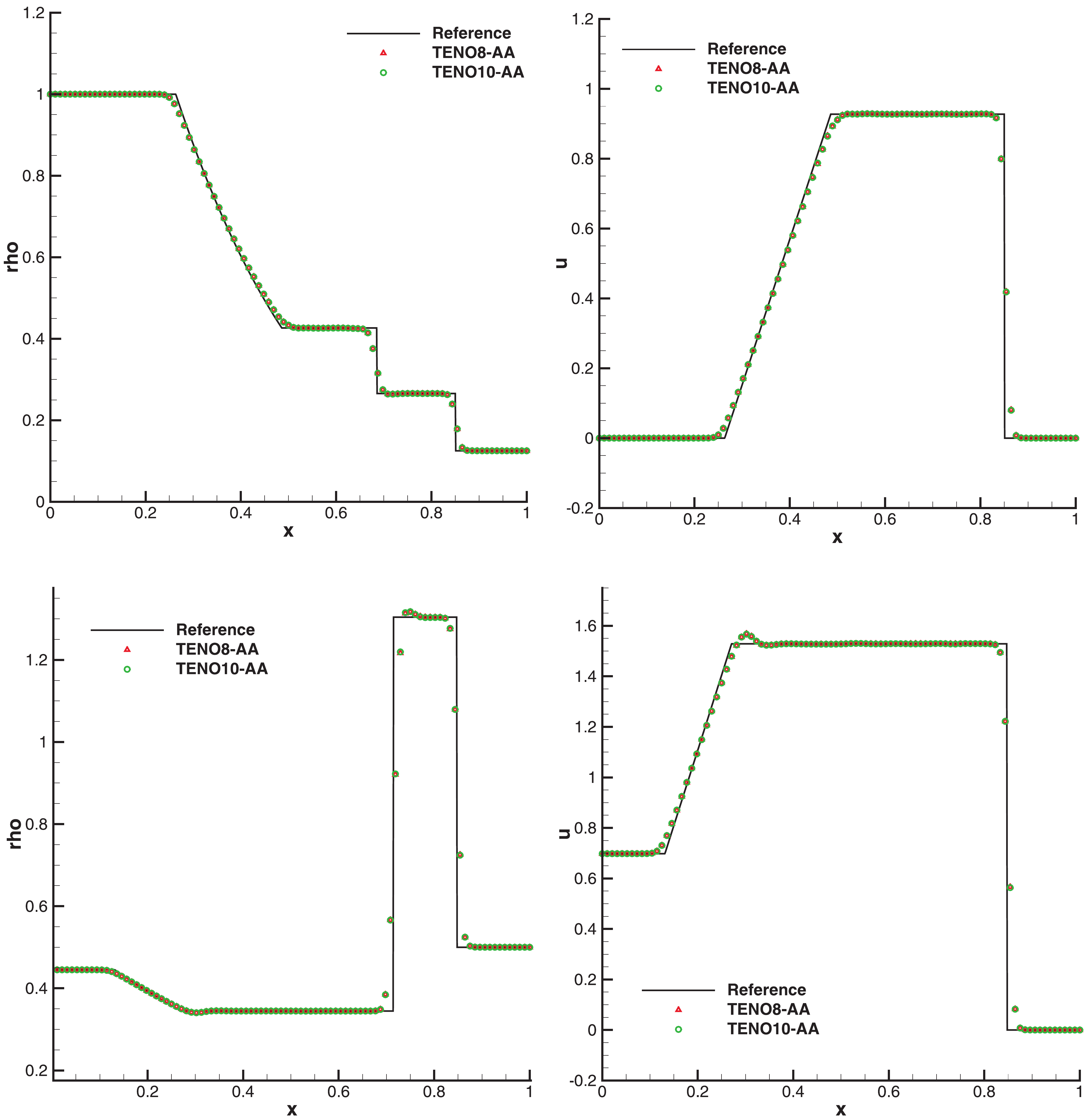}
\caption{Shock-tube simulations: the Sod's problem (top) and the Lax's problem (bottom). Left: density profiles; right: velocity profiles. Discretization on 96 uniformly distributed grid points. References represent the theoretical solutions of the Riemann problem. }
\label{Fig:lax-sod}
\end{figure}

As shown in Fig.~\ref{Fig:lax-sod}, both the TENO8-AA and TENO10-AA schemes show sharp shock-capturing property and the computed solutions agree well with the references.

\subsection{Shock-density wave interaction}
The first case is proposed by Shu and Osher \cite{Shu1989}. A one-dimensional Mach 3 shock interacts with a perturbed density field generating both small-scale structures and discontinuities, and hence it is selected to validate shock-capturing and wave-resolution capability. The initial condition is given as
\begin{equation}
\label{eq:shuosher}
(\rho ,u,p) = \left\{ {\begin{array}{*{20}{c}}
{(3.857,2.629,10.333),}&{\text{if } 0 \le x < 1} ,\\
{(1 + 0.2\sin (5(x-5)),0,1),}&{\text{if } 1 \le x \le 10} .
\end{array}} \right.
\end{equation}
The computational domain is [0,10] with $N = 200$ uniformly distributed mesh cells and the final evolution time is $t = 1.8$. The reference solution is obtained by the fifth-order WENO5-JS scheme with $N = 2000$.

As an extension of the Shu-Osher problem, the second case is proposed by Titarev and Toro \cite{Titarev2004} to test a severely oscillatory wave interacting with shock discontinuity. The initial condition is given as
\begin{equation}
\label{eq:Toro}
(\rho ,u,p) = \left\{ {\begin{array}{*{20}{c}}
{(1.515695,0.523346,1.805),}&{\text{if } 0 \le x < 0.5,}\\
{(1 + 0.1\sin (20\pi (x - 5)),0,1),}&{\text{if } 0.5 \le x \le 10.}
\end{array}} \right.
\end{equation}
The computational domain is [0,10] with $N = 1000$ uniformly distributed mesh cells and the final evolution time is $t = 5$. The reference solution is obtained by the fifth-order WENO5-JS scheme with $N = 5000$. A CFL number of 0.1 is employed.

\begin{figure}%
\centering
  \includegraphics[width=0.9\textwidth]{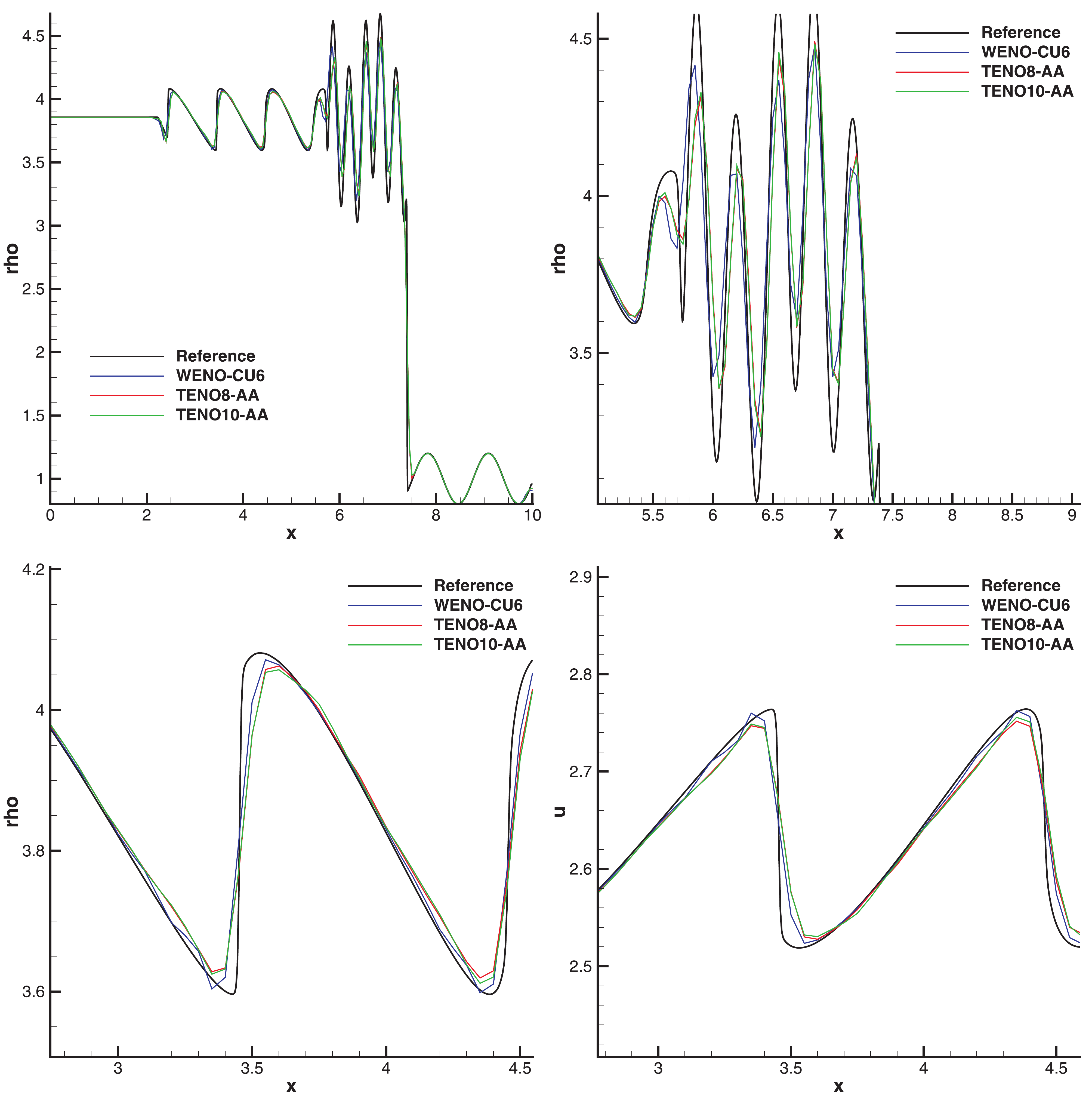}
\caption{Shu-Osher shock-density wave interaction problem: solutions from the TENO8-AA and TENO10-AA schemes. Top: density distribution (left) and a zoomed-in view of the density distribution (right). Bottom: zoomed-in views of the density profile (left) and the velocity profile (right). Discretization on 200 uniformly distributed grid points. }
 \label{Fig:shu-osher}
\end{figure}

As shown in Fig.~\ref{Fig:shu-osher}, the TENO8-AA and TENO10-AA schemes perform better in resolving the high-wavenumber density fluctuations. Furthermore, while WENO-CU6 generates numerical density and velocity oscillations in the vicinity of shocklets, the numerical solutions from TENO8-AA and TENO10-AA are sharp.

\begin{figure}%
\centering
  \includegraphics[width=0.9\textwidth]{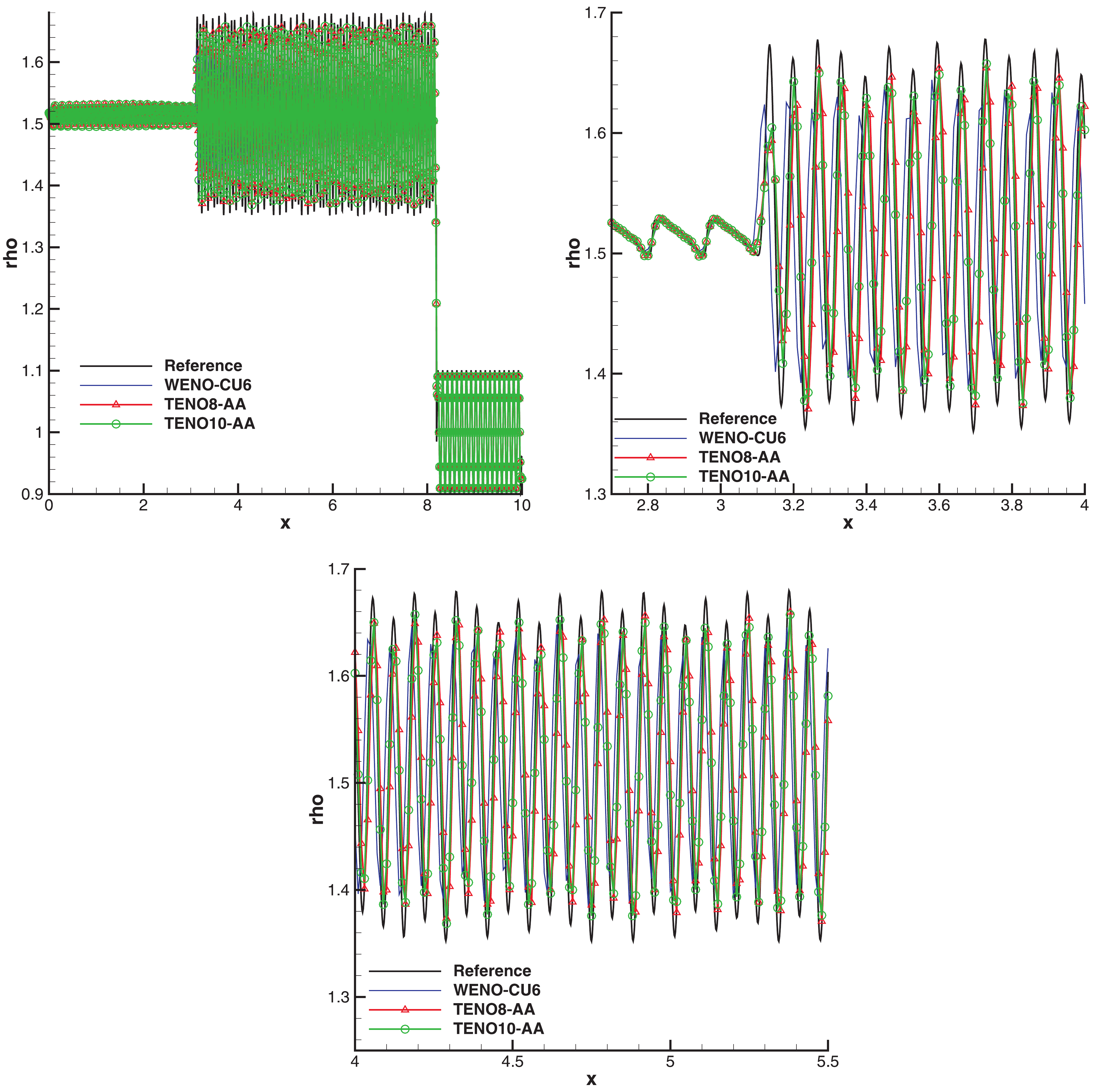}
\caption{Titarev-Toro shock-density wave interaction problem: solutions from the TENO8-AA and TENO10-AA schemes. Top: density distribution (left) and a zoomed-in view of the density distribution (right). Bottom: a zoomed-in view of the density profile. Discretization on 1000 uniformly distributed grid points. }
 \label{Fig:Titarev-Toro}
\end{figure}

As shown in Fig.~\ref{Fig:Titarev-Toro}, the results from TENO8-AA and TENO10-AA schemes agree with the reference significantly better than WENO-CU6 in terms of both the density-wave amplitude and the phase.
\subsection{Interacting blast waves}
The two-blast-wave interaction taken from Woodward and Colella \cite{Woodward1984} is considered. The initial condition is
\begin{equation}
\label{eq:blastwaves}
(\rho ,u,p) = \left\{ {\begin{array}{*{20}{c}}
{(1,0,1000),}&{\text{if }0 \le x < 0.1},\\
{(1,0,0.01),}&{\text{if }0.1 \le x < 0.9},\\
{(1,0,100),}&{\text{if }0.9 \le x \le 1} .
\end{array}} \right.
\end{equation}
A reflective boundary condition is imposed at $x = 0$ and $x = 1$. The simulation is performed on a uniform mesh with $N = 400$ until the final time $t = 0.038$. The reference solution is produced by the fifth-order WENO5-JS scheme on a uniform mesh with $N = 2500$. The Roe scheme with entropy-fix is applied for evaluating the numerical flux.

\begin{figure}%
\centering
  \includegraphics[width=0.9\textwidth]{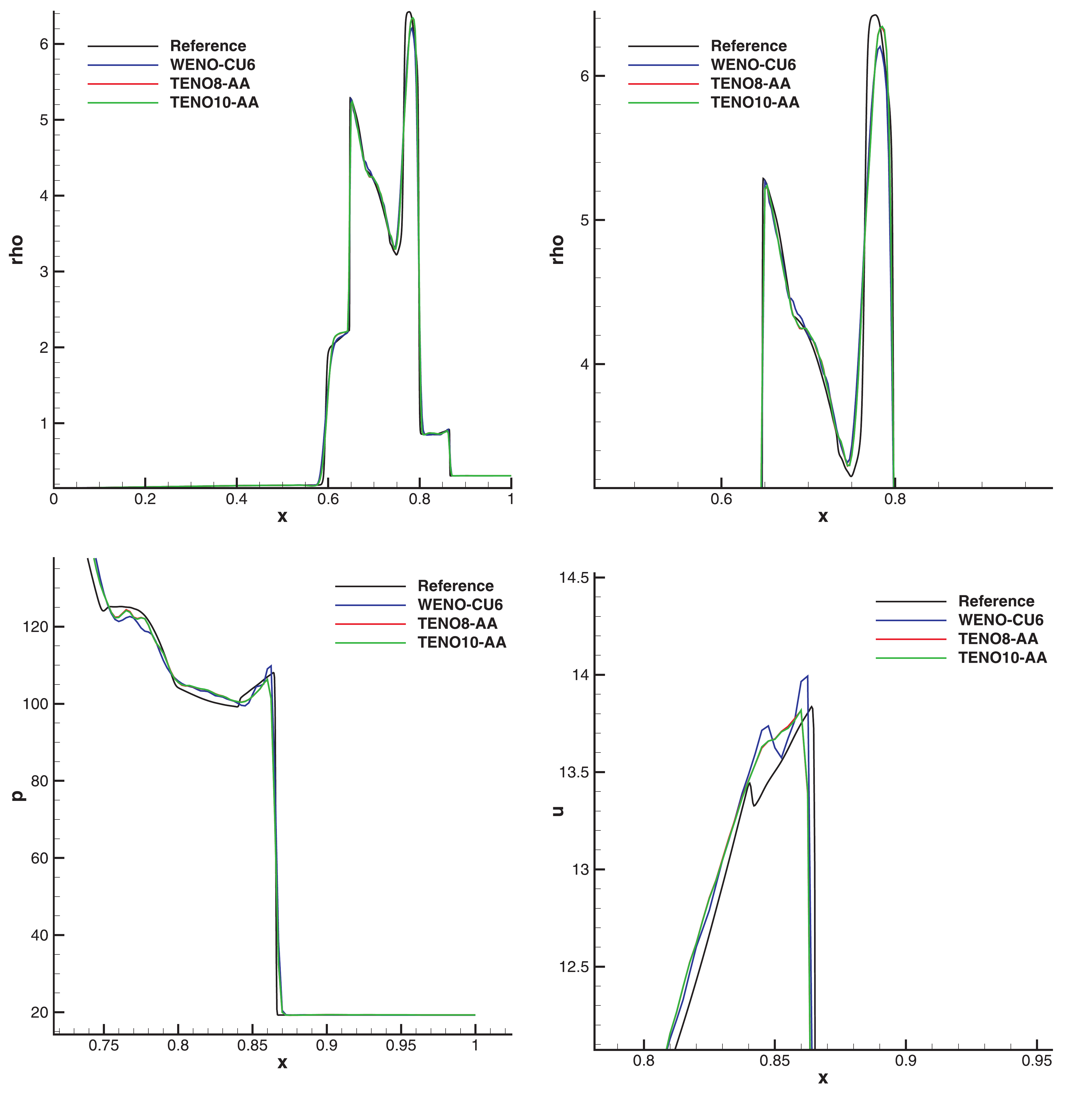}
\caption{Interacting blast waves problem: solutions from the TENO8-AA and TENO10-AA schemes. Top: density distribution (left) and a zoomed-in view of the density profile (right).  Bottom: a zoomed-in view of the pressure profile (left) and a zoomed-in view of the velocity profile (right). Discretization on 400 uniform grid points. }
\label{Fig:blastwaves}
\end{figure}

As shown in Fig.~\ref{Fig:blastwaves}, the TENO8-AA and TENO10-AA schemes resolve better density peak than WENO-CU6 at $x \approx 0.78$. Moreover, the results from TENO8-AA and TENO10-AA are free from the numerical oscillations in the pressure distribution and the overshoots in the velocity profile at $x \approx 0.86$, which are generated with WENO-CU6.

\subsubsection{Double Mach reflection of a strong shock}

The initial condition is \cite{Woodward1984}
\begin{equation}
\label{eq:DMR}
(\rho ,u,v,p) = \left\{ {\begin{array}{*{20}{c}}
{(1.4,0,0,1) ,}&{\text{if } y < 1.732(x - 0.1667) ,}\\
{(8,7.145, - 4.125,116.8333) , }&{\text{otherwise }.}
\end{array}} \right.
\end{equation}
The computational domain is $[0,4] \times [0,1]$ and the simulation end time is $t = 0.2$. Initially, a right-moving Mach 10 shock wave is placed at $x = 0.1667$ with an incident angle of ${60^0}$ to the x-axis. The post-shock condition is imposed from $x = 0$ to $x = 0.1667$ whereas a reflecting wall condition is enforced from $x = 0.1667$ to $x = 4$ at the bottom boundary. Other settings are the same with those in \cite{fu2016family}.

\begin{figure}%
\centering
\includegraphics[width=0.48\textwidth]{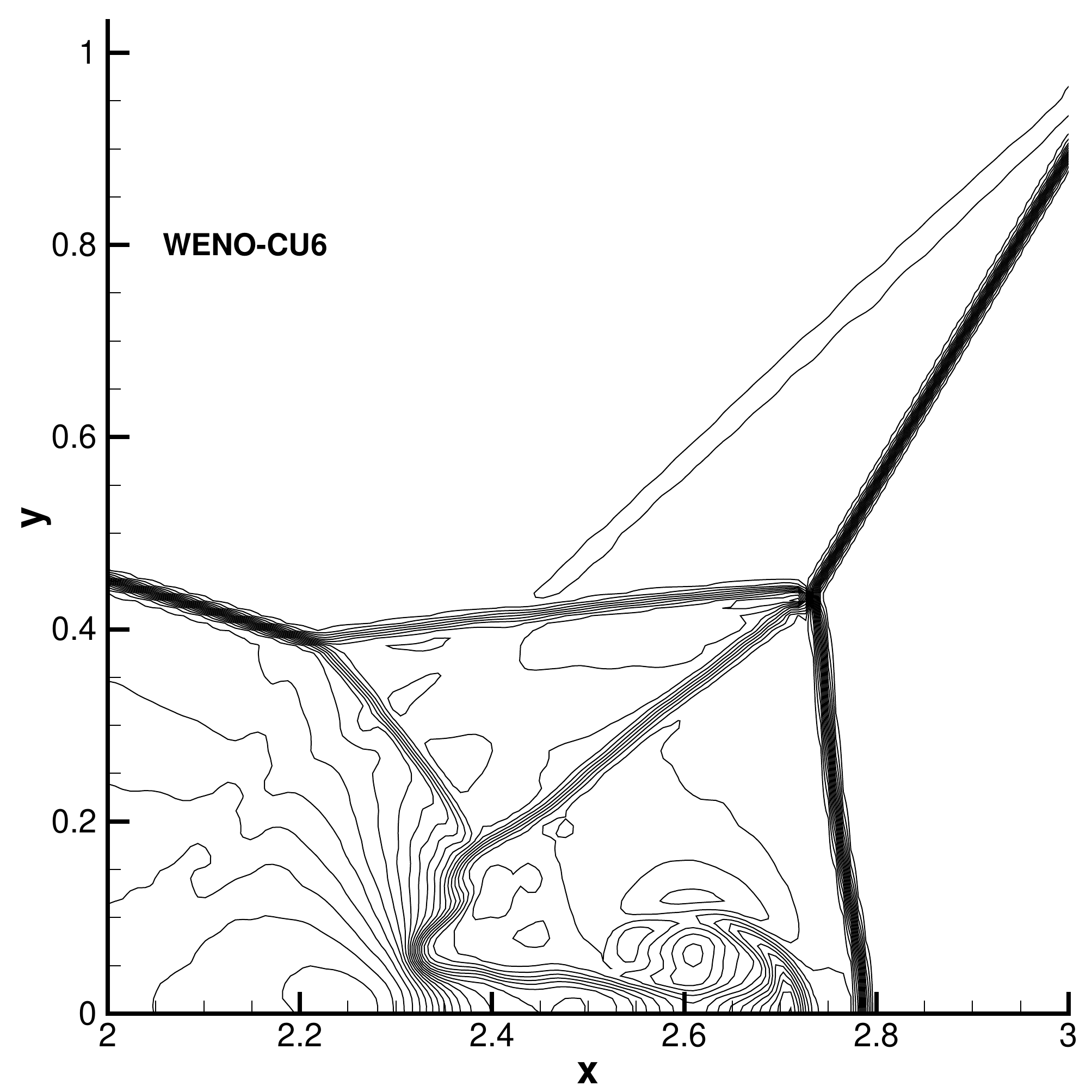}
\includegraphics[width=0.48\textwidth]{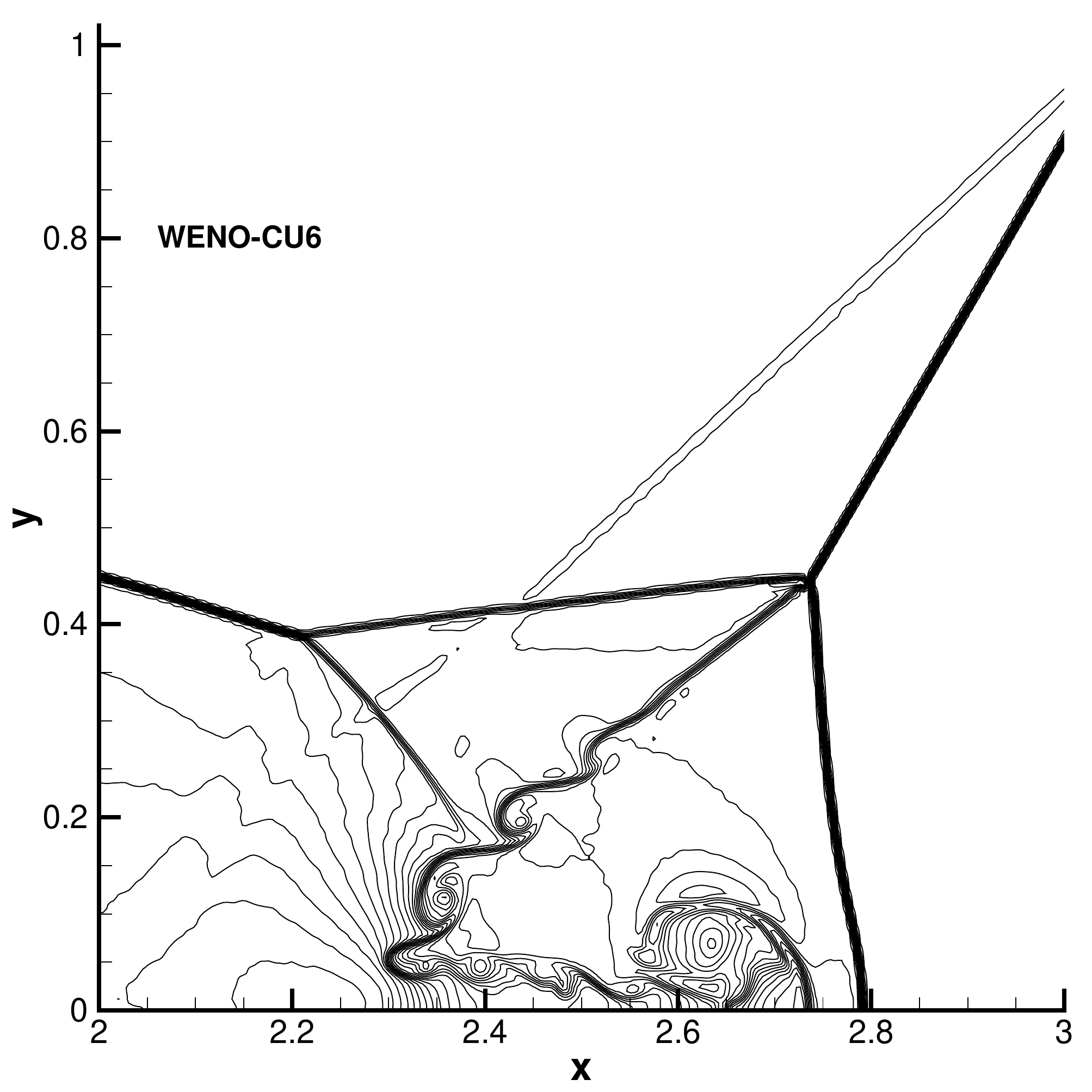}
  \caption{Double Mach reflection of a strong shock: density contours from the WENO-CU6 scheme at simulation time $t = 0.2$. Resolution of $512 \times 128$ (left) and $1024 \times 256$ (right). This figure is drawn with 43 density contours between 1.887 and 20.9.}
 \label{Fig:dmr_reference}
\end{figure}

\begin{figure}%
\centering
\includegraphics[width=0.48\textwidth]{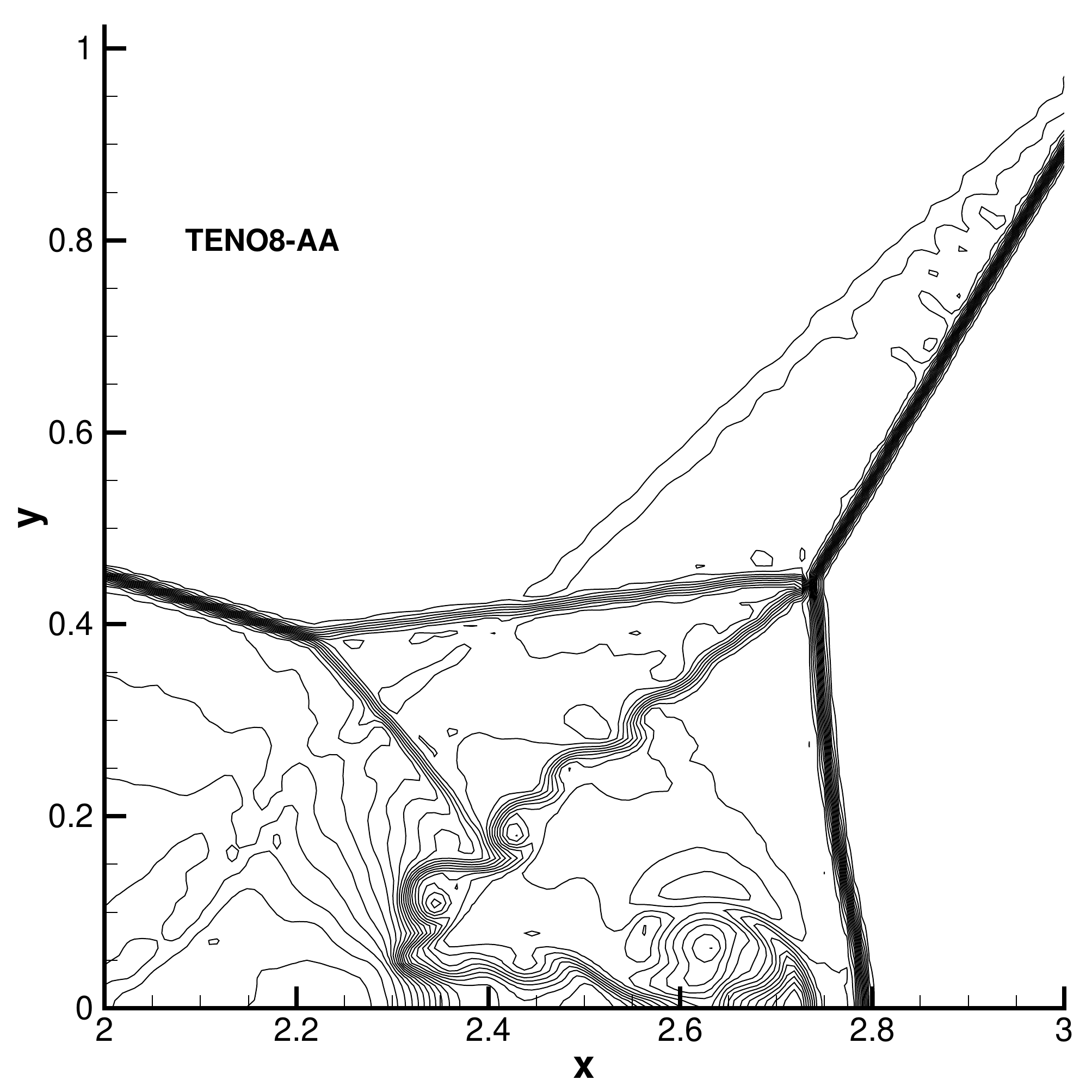}
\includegraphics[width=0.48\textwidth]{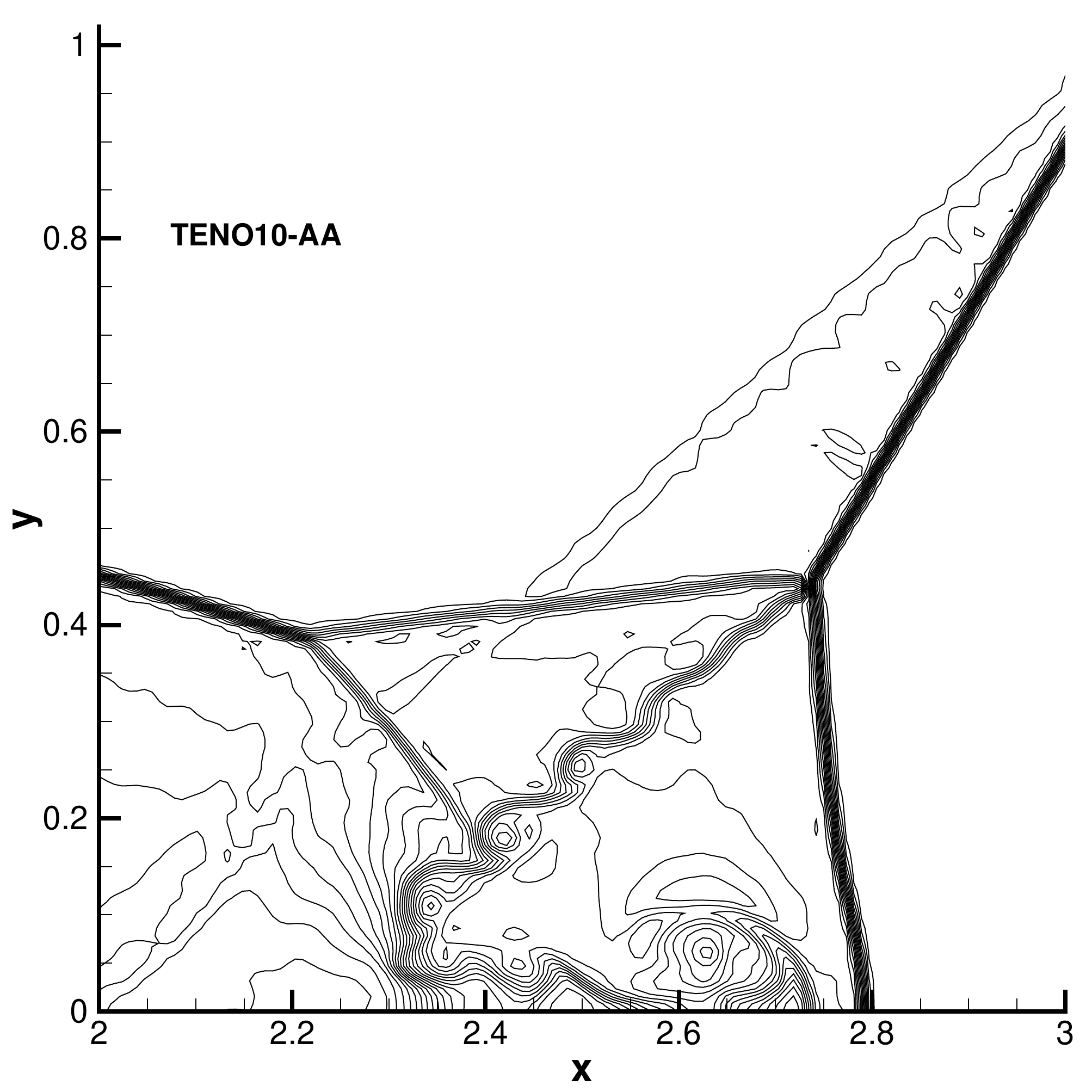}
\includegraphics[width=0.48\textwidth]{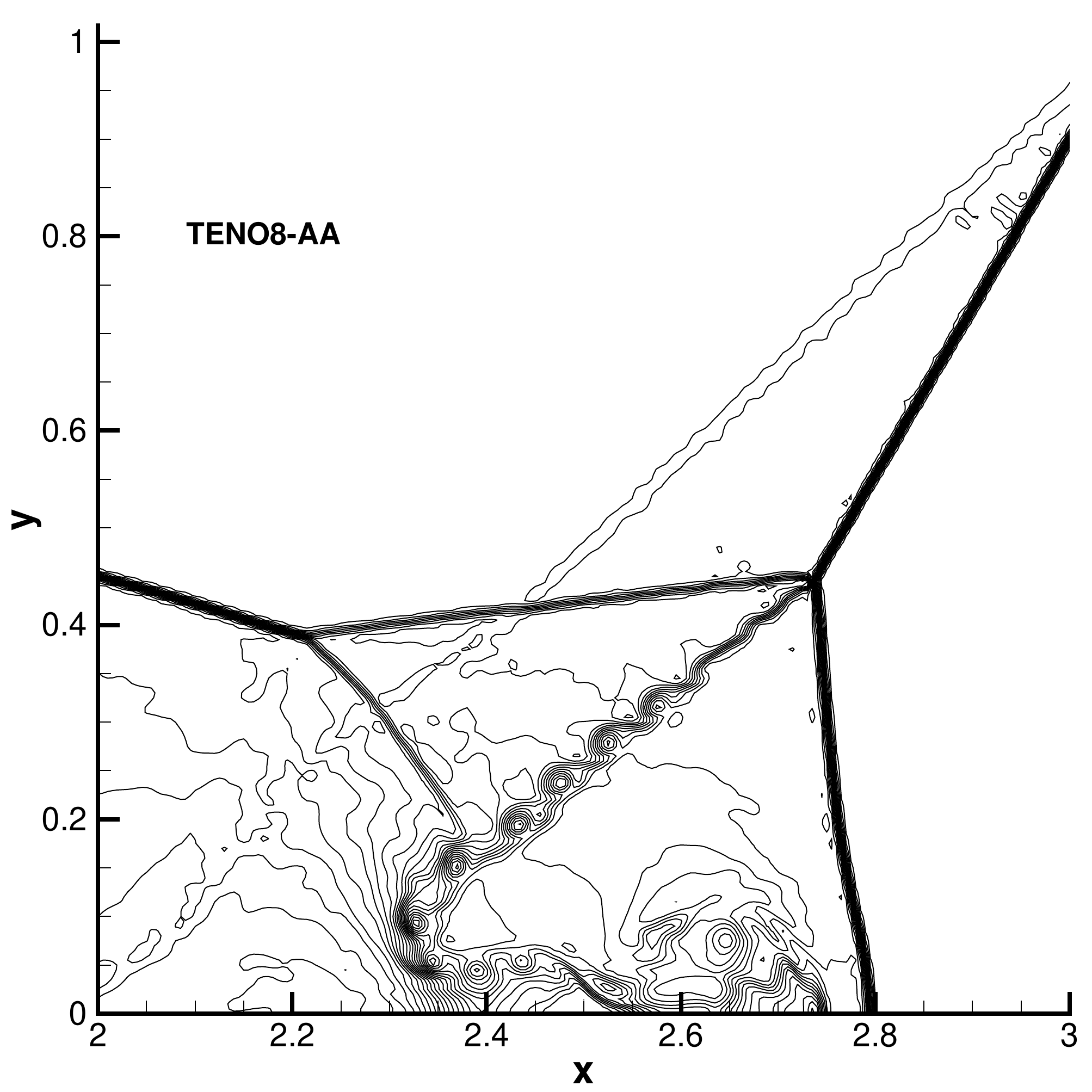}
\includegraphics[width=0.48\textwidth]{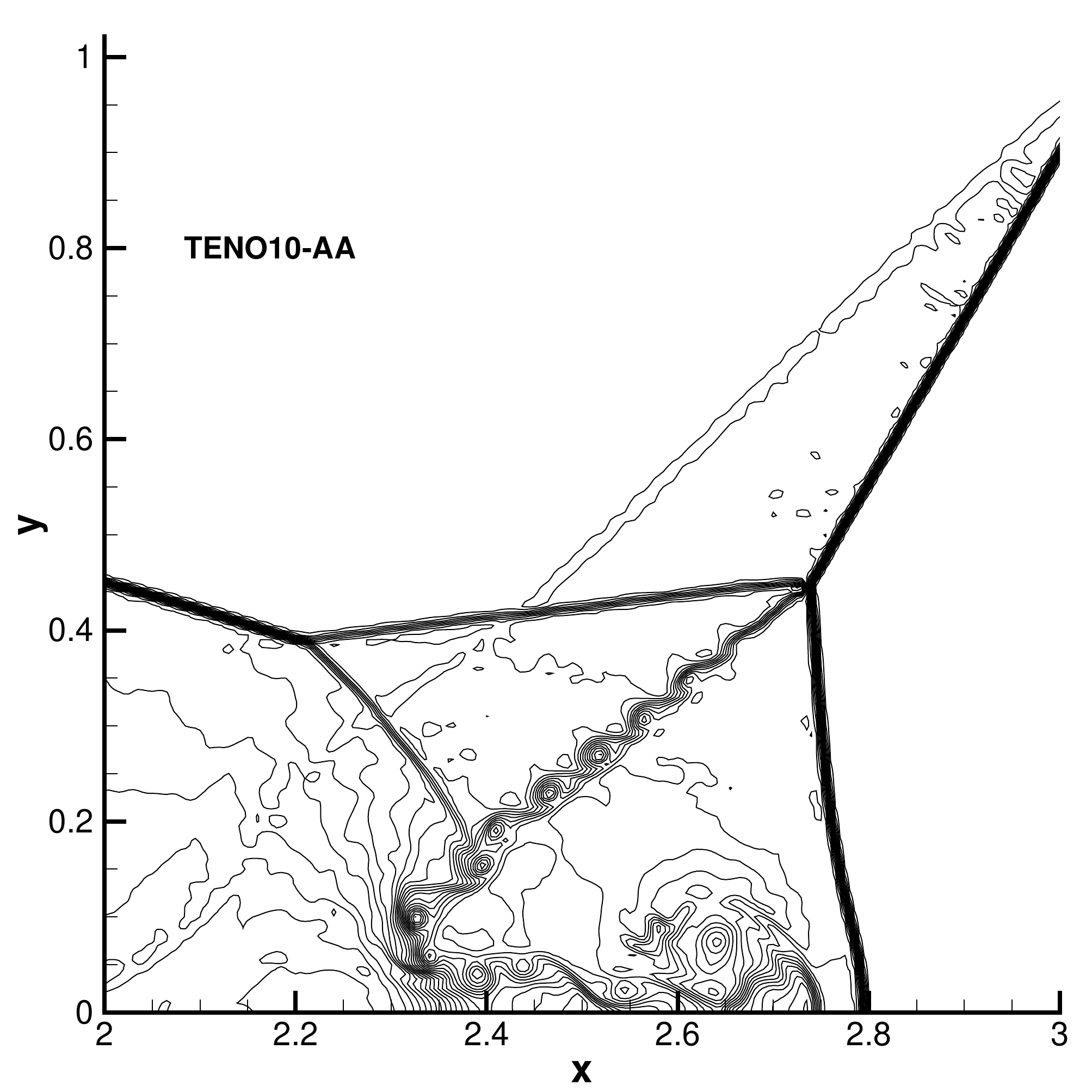}
  \caption{Double Mach reflection of a strong shock: density contours from the TENO8-AA and TENO10-AA schemes at simulation time $t = 0.2$. Resolution of $512 \times 128$ (top) and $800 \times 200$ (bottom). This figure is drawn with 43 density contours between 1.887 and 20.9.}
 \label{Fig:dmr_TENO-AA}
\end{figure}

The results from the WENO-CU6 scheme at resolution of $512 \times 128$ and $1024 \times 256$ are given in Fig.~\ref{Fig:dmr_reference} for comparison.
As shown in Fig.~\ref{Fig:dmr_TENO-AA}, the proposed TENO8-AA and TENO10-AA scheme perform much better than WENO-CU6. With the resolution of $512 \times 128$, TENO8-AA and TENO10-AA resolve much more small-scale vortical structures than WENO-CU6 with a resolution of $1024 \times 256$. The present results represent the best among those from other low-dissipation schemes at similar resolutions, see Fig.~17 and Fig.~18 of \cite{fu2016family}.
\subsubsection{Rayleigh-Taylor instability}
The initial condition is \cite{Xu2005}
\begin{equation}
(\rho ,u,v,p) = \left\{ {\begin{array}{*{20}{c}}
{(2,0, - 0.025c\cos (8\pi x),1 + 2y) ,}&{\text{if } 0 \le y < 1/2 ,}\\
{(1,0, - 0.025c\cos (8\pi x),y + 3/2) ,}&{\text{if }1/2 \le y \le 1 .}
\end{array}} \right.
\end{equation}
where the sound speed $c = \sqrt {\gamma \frac{p}{\rho }}$ with $\gamma  = \frac{5}{3}$. The computational domain is $[0,0.25] \times [0,1]$. Reflective boundary conditions are imposed at the left and right sides of the domain. Constant primitive variables $(\rho ,u,v,p) = (2,0,0,1)$ and $(\rho ,u,v,p) = (1,0,0,2.5)$ are applied for the bottom and top boundary, respectively.

\begin{figure}%
\centering
\includegraphics[width=0.18\textwidth]{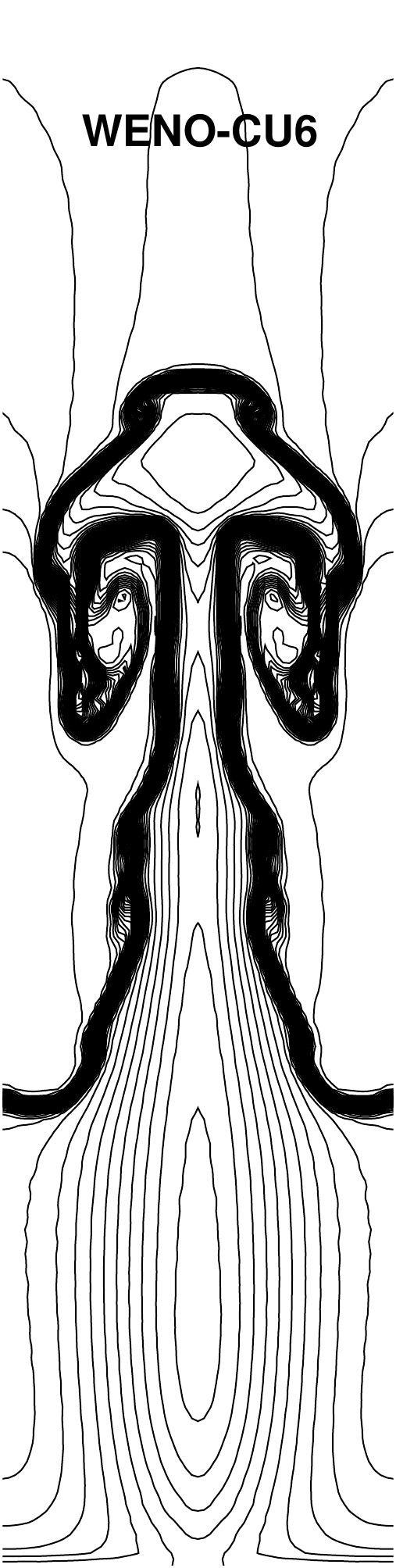}
\includegraphics[width=0.18\textwidth]{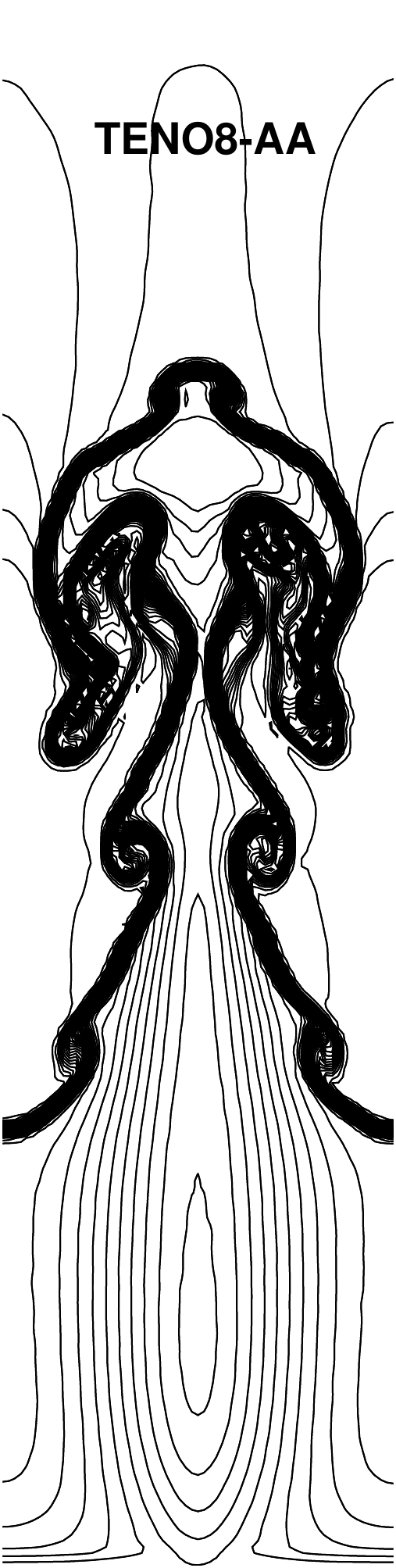}
\includegraphics[width=0.18\textwidth]{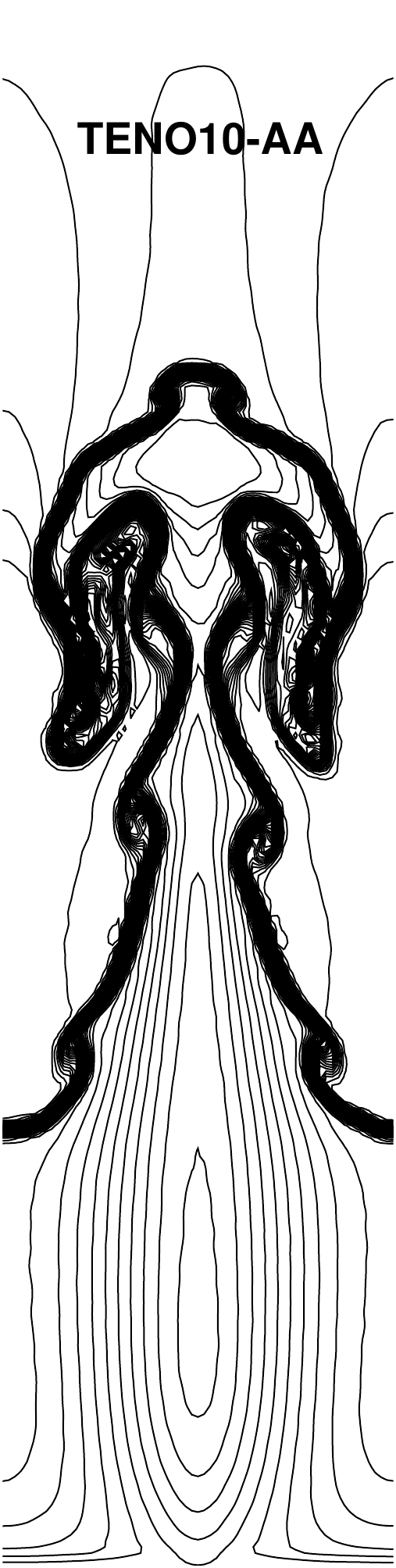} \\
\includegraphics[width=0.18\textwidth]{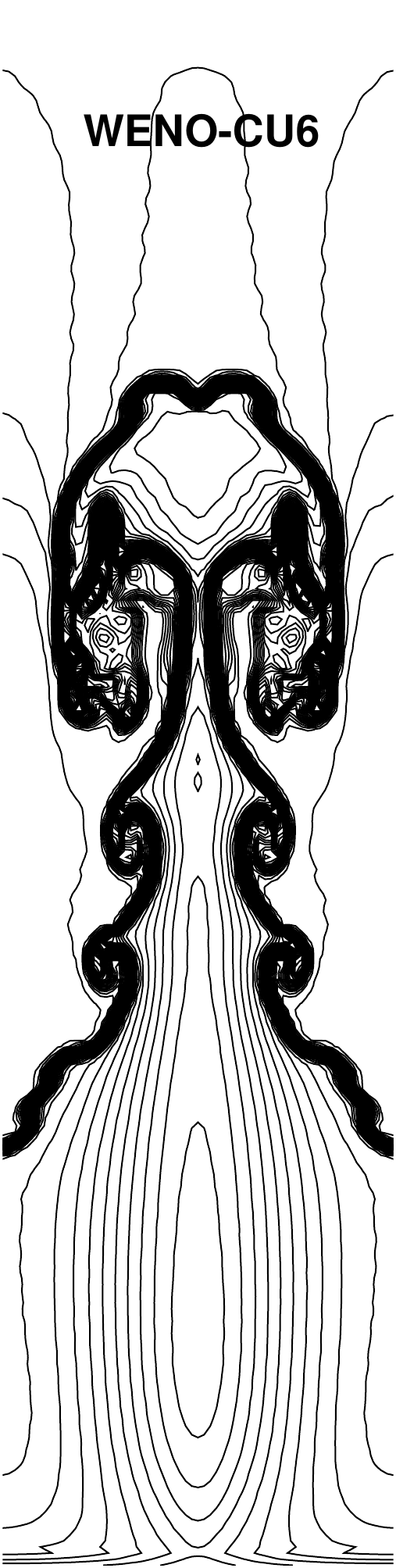}
\includegraphics[width=0.18\textwidth]{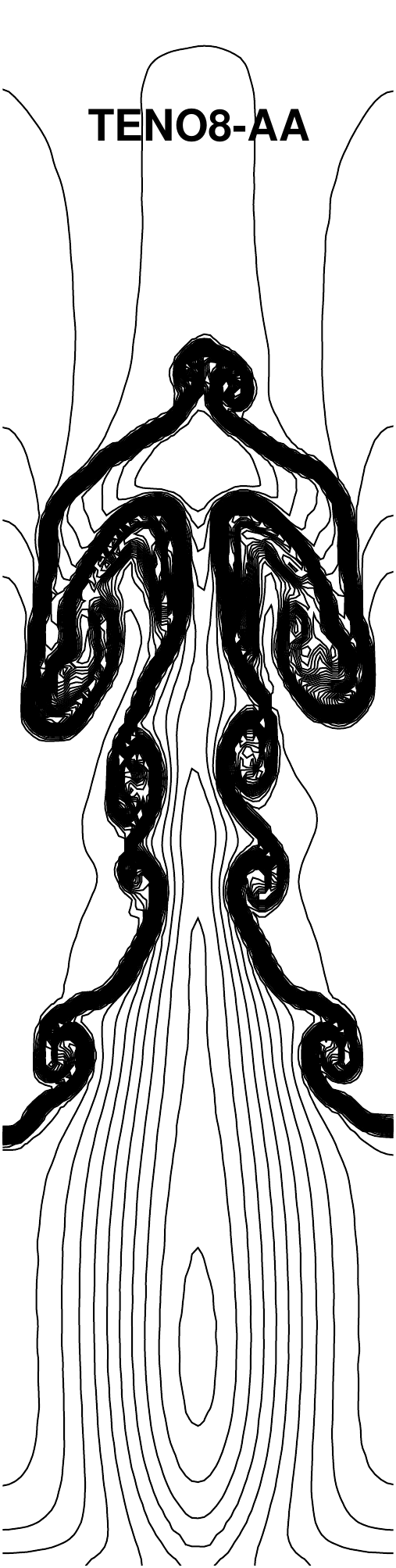}
\includegraphics[width=0.18\textwidth]{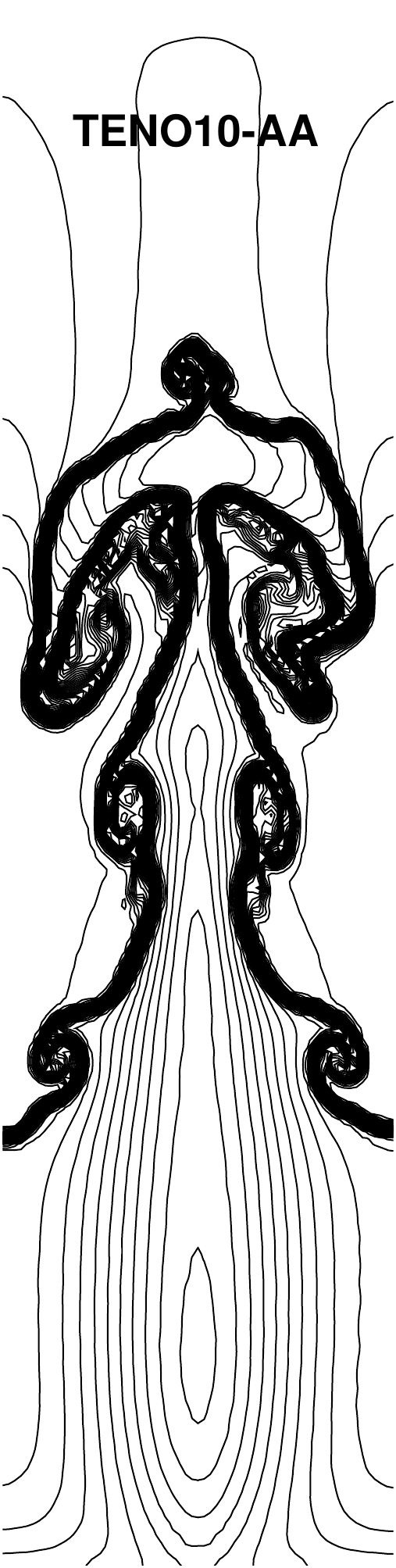}
  \caption{Rayleigh-Taylor instability: density contours from the TENO8-AA and TENO10-AA schemes with a resolution of $64 \times 256$. Flux computing with the Rusanov (top) and Roe (bottom) scheme. The simulation is run until $t = 1.95$. This figure is drawn with 43 density contours between 0.9 and 2.2.}
 \label{Fig:RTI}
\end{figure}

As shown in Fig.~\ref{Fig:RTI}, the present TENO8-AA and TENO10-AA schemes resolve more fine structures than WENO-CU6, indicating much less numerical dissipation. 
{It is also observed that the results from the present schemes lose the flow symmetry. 
For Rayleigh-Taylor instability, results from low-dissipation shock-capturing schemes may lose symmetry, see e.g., the result from WENO-SYMBO (Fig.13 of \cite{sun2014sixth}), WENO-CU6 (Fig.20 of \cite{fu2016family} with higher resolution), and WENO9 (Fig.2 of \cite{shi2003resolution}). The main reason is that the machine round-off errors, which tend to break the symmetry, cannot be hidden by the low numerical dissipation of the present schemes \cite{fu2016family}\cite{fu2017targeted}\cite{fu2018new}\cite{fleischmann2019numerical}.  It has been recently shown that careful floating point arithmetic implementation \cite{fleischmann2019numerical} may dramatically reduce round-off errors so that the flow symmetry is maintained until very late stages even for inviscid physically unstable flows.}

{\subsection{2D Kelvin-Helmholtz instability (KHI)}
    The Kelvin-Helmholtz instability with single mode perturbation is considered \cite{san2015evaluation}\cite{ryu2000magnetohydrodynamic}. The computational domain is $[-0.5, 0.5] \times [-0.5, 0.5]$ and the initial conditions are given as
    \begin{equation}
        (\rho ,{u}) = \left\{ {\begin{array}{*{20}{c}}
        {(2, - 0.5),}&{\left| y \right| \le 0.25,}\\
        {(1,0.5),}&{\text{otherwise,}} \\
        \end{array}} \right. \\
    \end{equation}
    and
    \begin{equation}
        ({v},p) = (0.01\sin (2\pi x),2.5).
    \end{equation}
    $\gamma = 1.4$ and periodic boundary conditions are enforced at the boundaries of the computational domain. The final simulation time is $t=1$.

 As shown in Fig.~\ref{Fig:KHI}, at the coarse resolution of $512 \times 512$, the present TENO8-AA and TENO10-AA schemes exhibit lower numerical dissipation by resolving more small-scale structures than WENO-CU6.
\begin{figure}%
\centering
\includegraphics[width=0.45\textwidth]{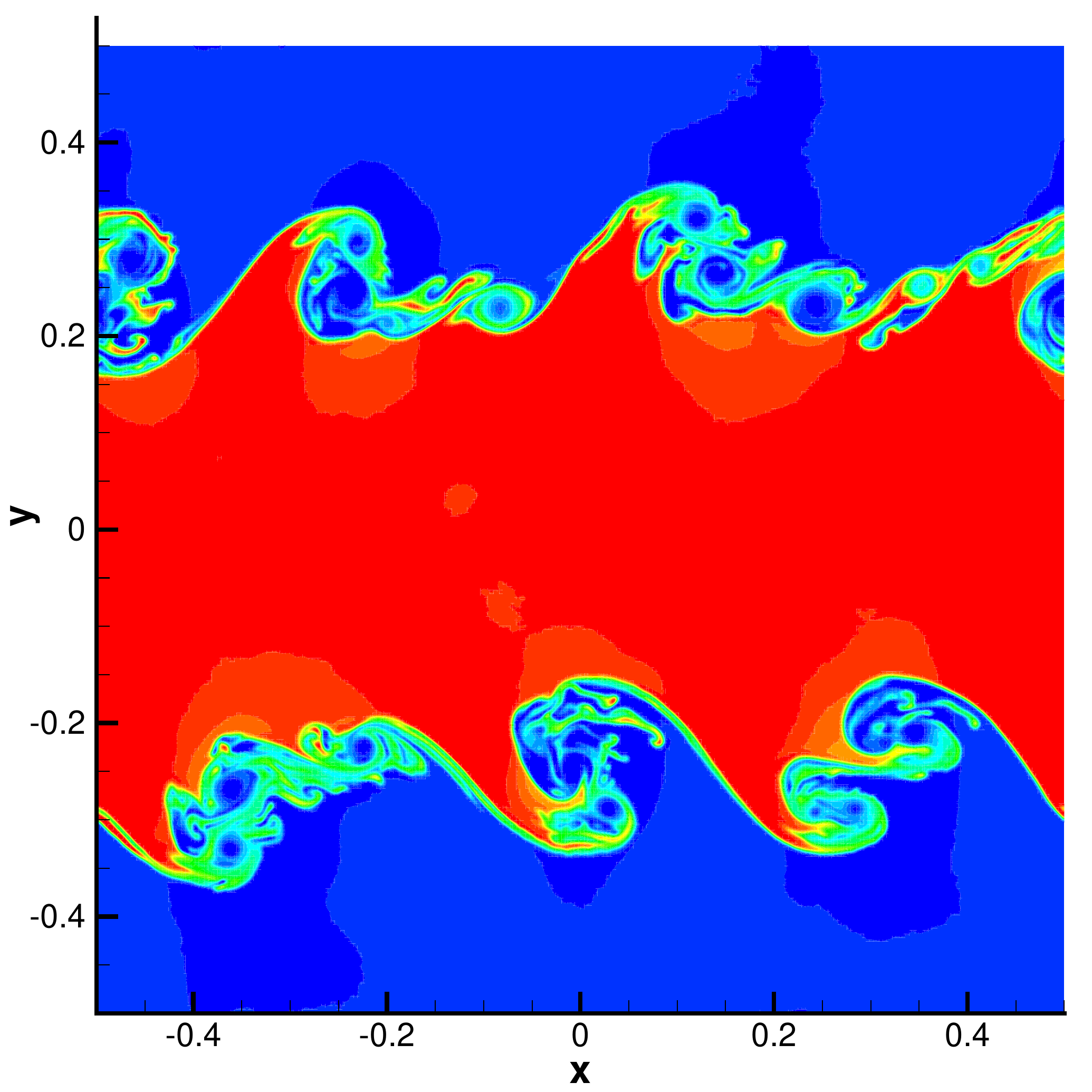} 
\includegraphics[width=0.45\textwidth]{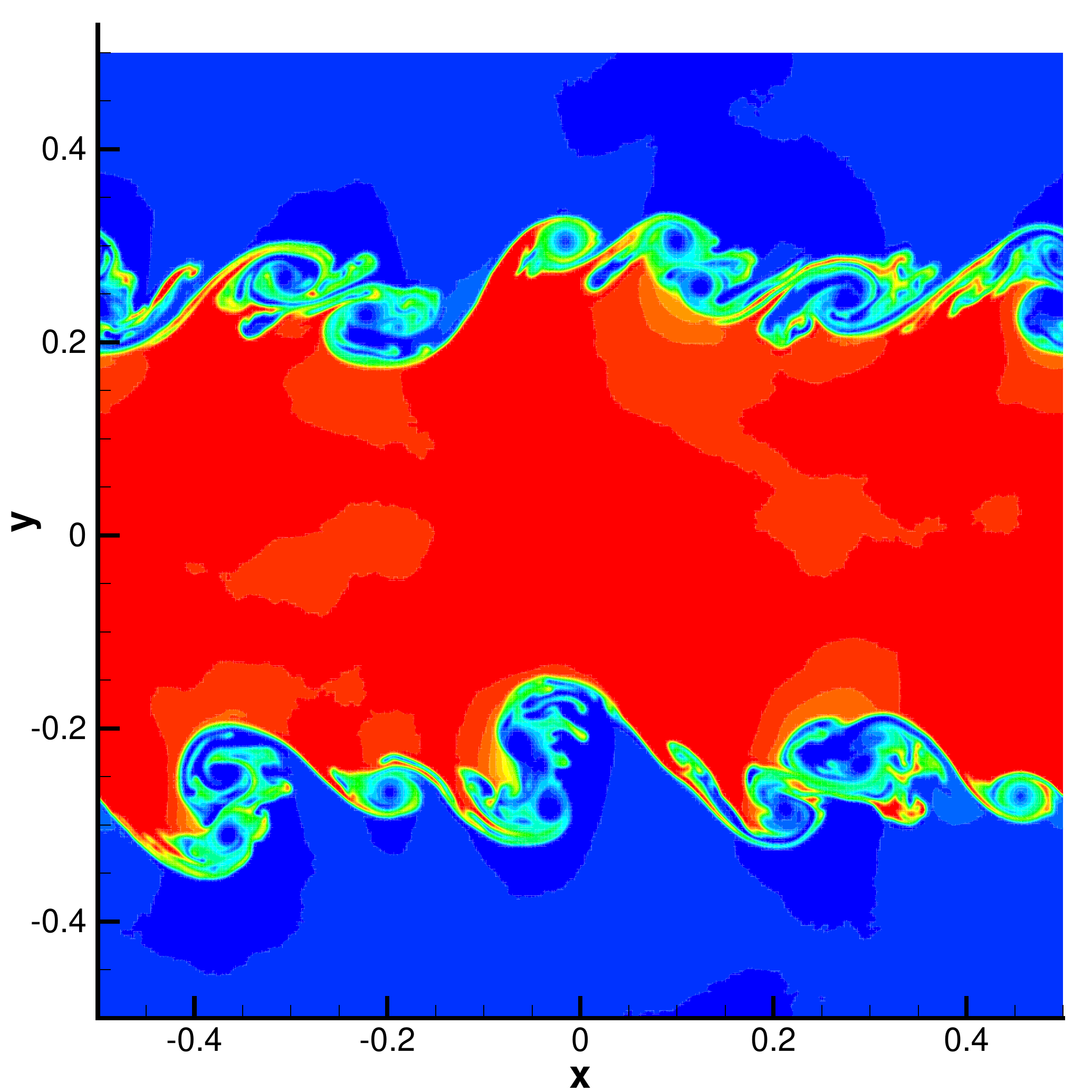} \\
\includegraphics[width=0.45\textwidth]{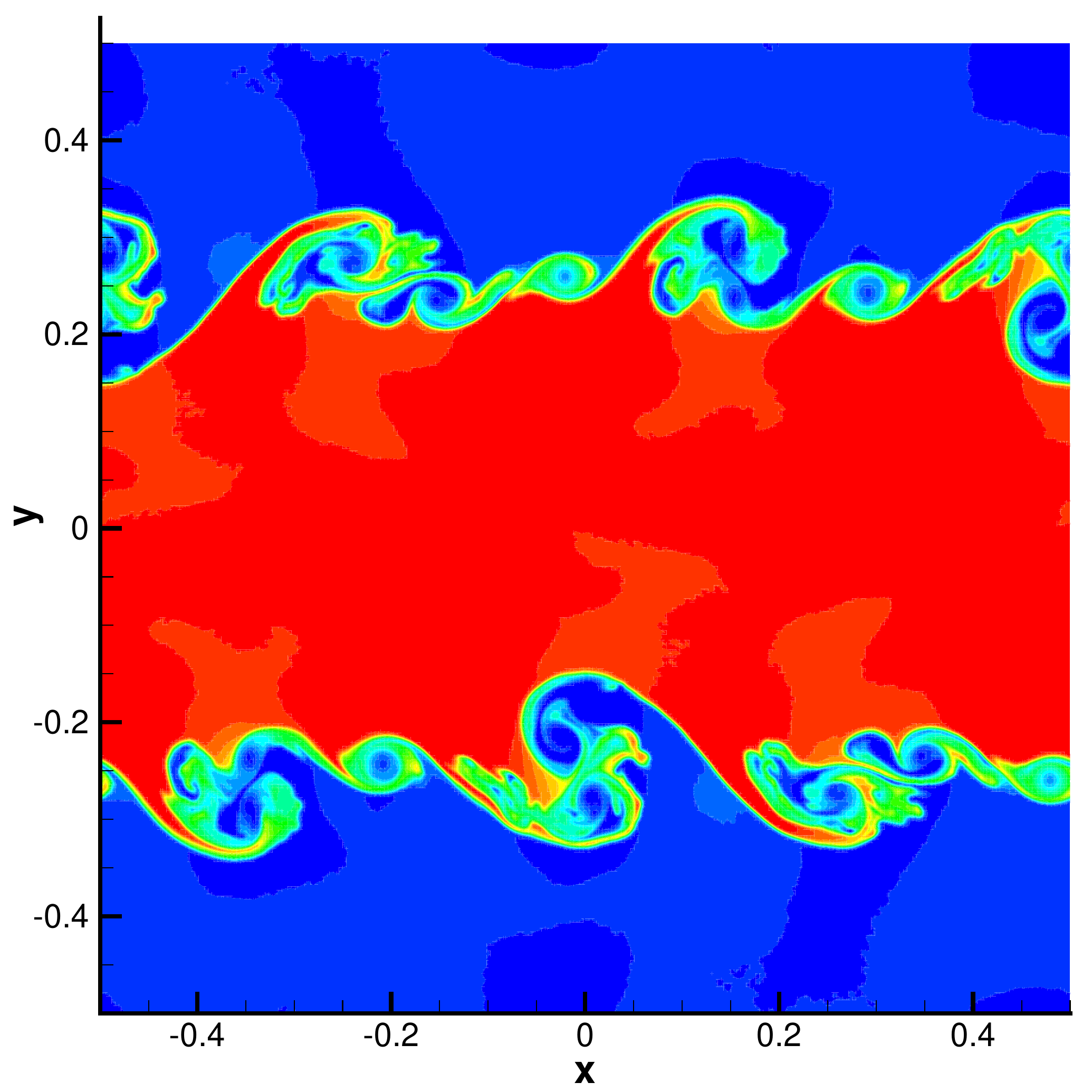}
  \caption{{2D Kelvin-Helmholtz instability: density contours from the TENO8-AA (top left), TENO10-AA (top right), and WENO-CU6 (bottom) schemes with a resolution of $512 \times 512$.}}
 \label{Fig:KHI}
\end{figure}

 } 

\subsection{Extreme simulations}

For this part, the LLF scheme \cite{Jiang1996} is adopted for flux splitting if not mentioned otherwise. In addition, the positivity-preserving flux limiter \cite{hu2013positivity} is applied for stability.

\subsubsection{1D low density and low pressure problems}

Two 1D test problems involving vacuum or strong discontinuity, i.e. the double rarefaction problem \cite{hu2013positivity} and the Le Blanc problem \cite{loubere2005subcell}, are considered. For the first problem, the initial condition is \cite{hu2013positivity}
 \begin{equation}
 \label{eq:blastwaves}
 (\rho ,u,p) = \left\{ {\begin{array}{*{20}{c}}
 {(1,-2,0.1),}&{\text{if }0 \le x < 0.5},\\
 {(1,2,0.1),}&{\text{if }0.5 \le x \le 1} .
 \end{array}} \right.
 \end{equation}
The final simulation time is $t = 0.1$ and the mesh resolution is $N = 400$. The reference solution is given by Riemann solver analytically.

For the second case, the initial condition is \cite{loubere2005subcell}
 \begin{equation}
 \label{eq:LeBlancblastwaves}
 (\rho ,u,p) = \left\{ {\begin{array}{*{20}{c}}
 {(1,0,\frac{2}{3} \times 10^{-1}),}&{\text{if }0 \le x < 3},\\
 {(10^{-3},0,\frac{2}{3} \times 10^{-10}),}&{\text{if }3 \le x \le 9} .
 \end{array}} \right.
 \end{equation}
$\gamma  = \frac{5}{3}$. The mesh resolution is $N = 800$ and the final simulation time is $t = 6$. The reference solution is computed with the fifth-order WENO5-JS scheme at resolution of $N = 4000$.

\begin{figure}%
\centering
\includegraphics[width=0.85\textwidth]{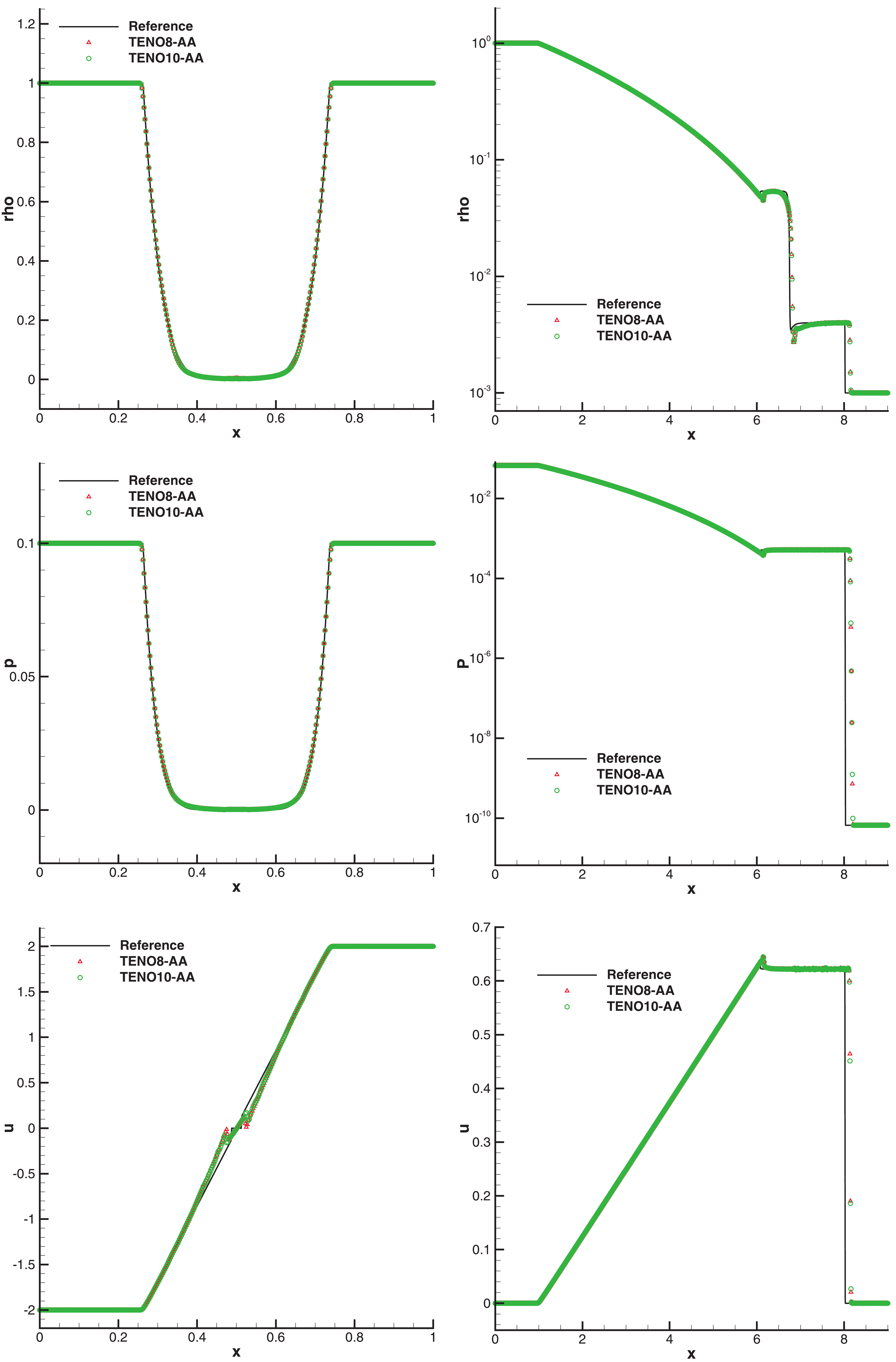}
  \caption{1D low density and low pressure problems: the double rarefaction problem (left) and the Le Blanc problem (right). Included are the distributions of flow density (top), pressure (middle) and velocity (bottom). }
 \label{Fig:1D_low_density_and_low_pressure_problems}
\end{figure}

As shown in Fig.\ref{Fig:1D_low_density_and_low_pressure_problems}, the proposed TENO8-AA and TENO10-AA schemes preserve good ENO property for discontinuity capturing and the computed results agree well with the reference solutions.

\subsubsection{2D low density and low pressure problem}

We consider the 2D Sedov blast wave problem involving low density and low pressure. The computational domain is $[0, 1.1] \times [0, 1.1]$. The initial condition is \cite{hu2013positivity}
 \begin{equation}
 \label{eq:blastwaves}
 (\rho , u, v, p) = \left\{ {\begin{array}{*{20}{c}}
  {(1, 0, 0, \frac{9.79264}{{\Delta x} { \Delta y}} \times 10^{4}),}&{\text{if } x \leq { \Delta x} \text{ and }  x \leq { \Delta y} } , \\
 {(1, 0, 0, 4 \times 10^{-13}),}&{\text{otherwise}} ,
 \end{array}} \right.
 \end{equation}
where $\Delta x = \Delta y = \frac{1.1}{160}$. The final simulation time is $t = 10^{-3}$. The symmetry condition is applied for the left and bottom boundaries while the outflow condition is employed for the other boundaries.

\begin{figure}%
\centering
\includegraphics[width=0.49\textwidth]{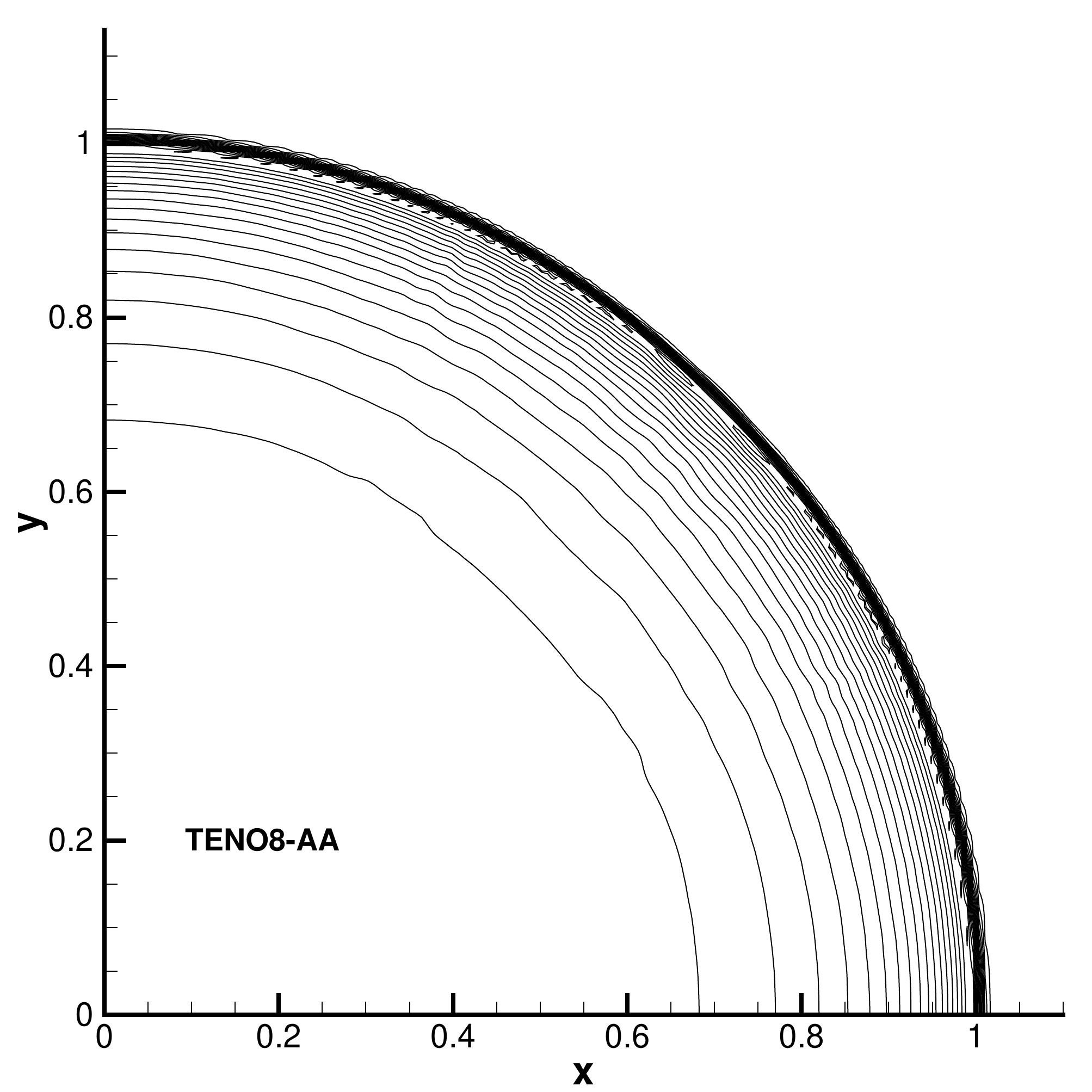}
\includegraphics[width=0.49\textwidth]{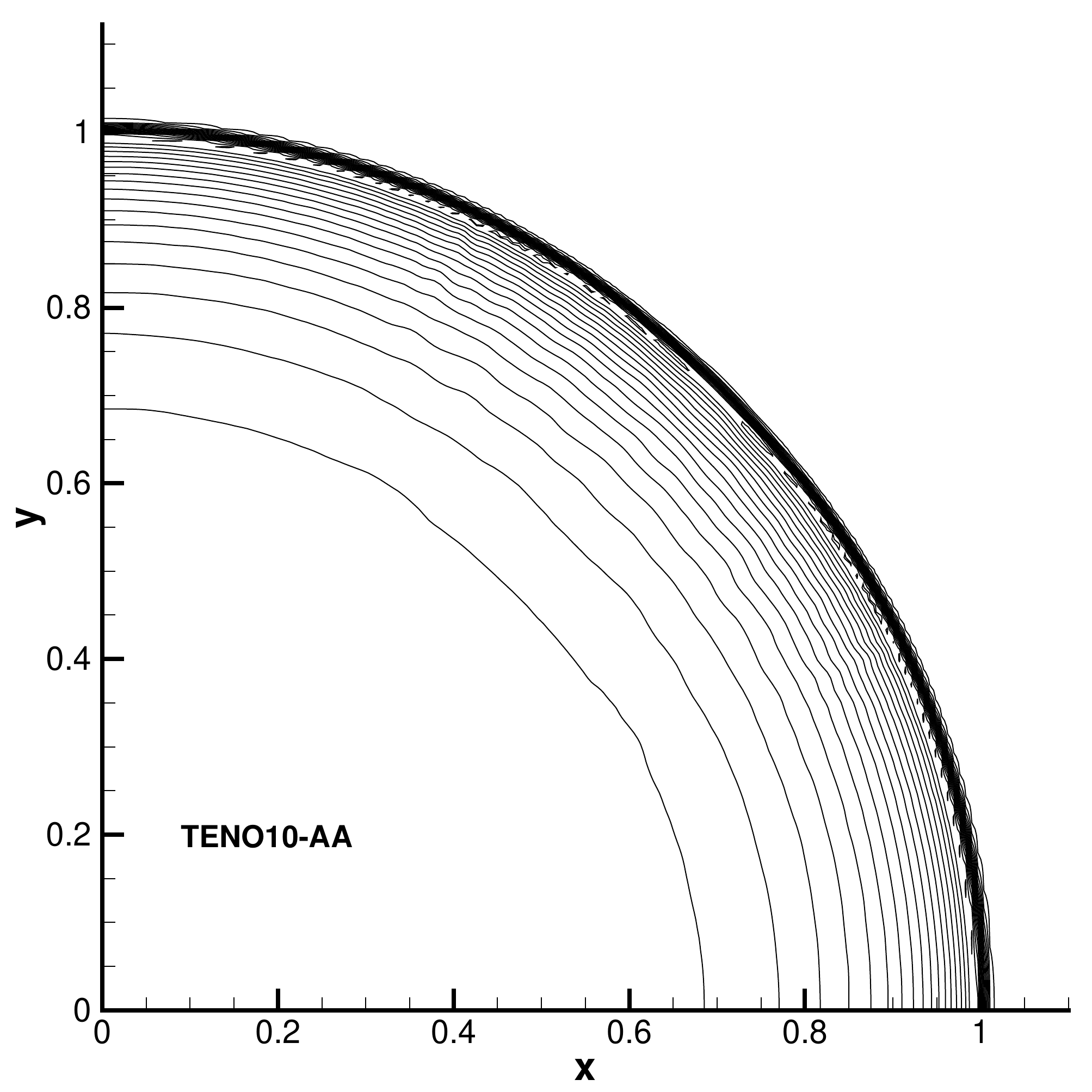} \\
\includegraphics[width=0.49\textwidth]{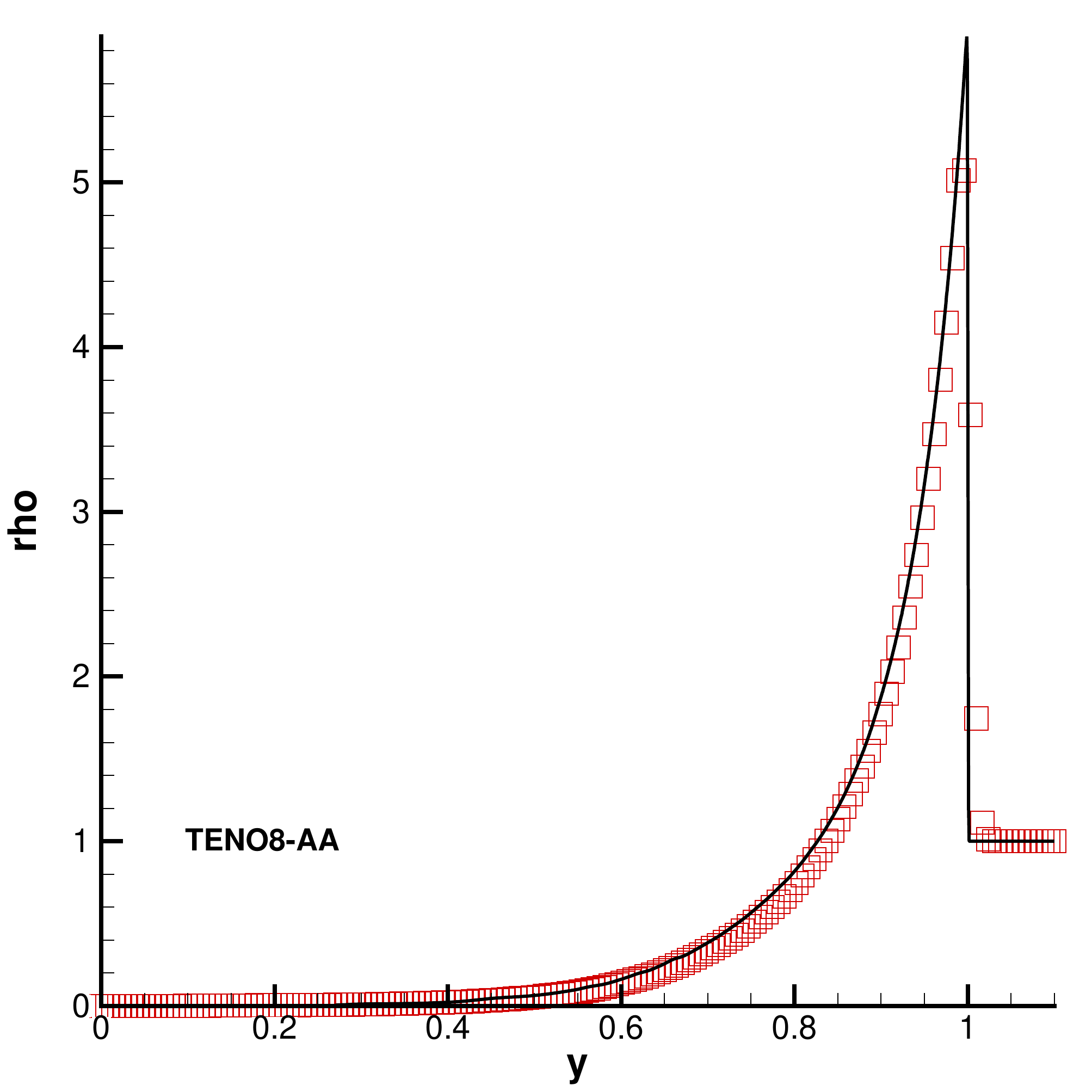}
\includegraphics[width=0.49\textwidth]{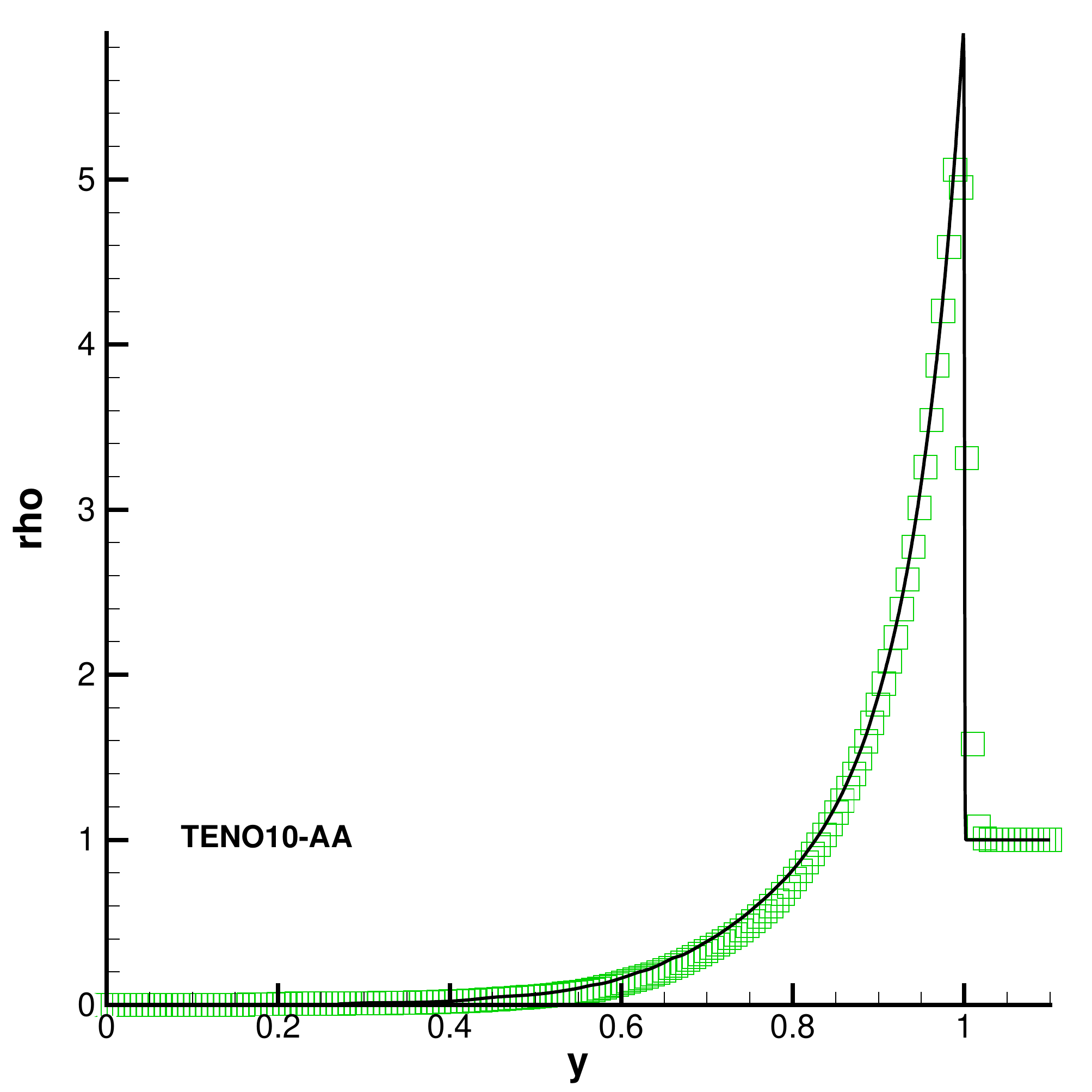}
  \caption{2D low density and low pressure problem: the computed results from TENO8-AA (left) and TENO10-AA (right) at a resolution of $160 \times 160$. Top: 20 equally-spaced density contourlines from 0 to 5.5; bottom: density profiles, where the solid line denotes the reference solution and the symbols denote the computed solution. }
 \label{Fig:2D_low-density_and_low-pressure_problems}
\end{figure}

As shown in Fig.~\ref{Fig:2D_low-density_and_low-pressure_problems}, both the TENO8-AA and TENO10-AA schemes capture the density discontinuity sharply and the resolved profiles agree well with the reference solution.
\subsubsection{2D high Mach number astrophysical jet}

The simulations of two 2D high Mach number astrophysical jets without radiative cooling are considered \cite{zhang2010positivity}. For the first case of Mach number 80, the computational domain is $[0 , 2] \times [-0.5 , 0.5]$ with full ambient gas of $(\rho,u,v,p) = (0.5, 0, 0, 0.4127)$. For the left boundary, $(\rho,u,v,p) = (5, 30, 0, 0.4127)$ if $y \in [-0.05 , 0.05]$ and $(\rho,u,v,p) = (5, 0, 0, 0.4127)$ otherwise. For the second case of Mach number 2000, the computational domain is $[0 , 1] \times [-0.25 , 0.25]$ with full ambient gas of $(\rho,u,v,p) = (0.5, 0, 0, 0.4127)$. For the left boundary, $(\rho,u,v,p) = (5, 800, 0, 0.4127)$ if $y \in [-0.05 , 0.05]$ and $(\rho,u,v,p) = (5, 0, 0, 0.4127)$ otherwise. For both cases, the right, top and bottom boundaries are imposed with the outflow condition and the ratio of the specific heats $\gamma  = \frac{5}{3}$. The mesh resolutions are $448 \times 224$ and $800 \times 400$ for the first and second cases, respectively.

\begin{figure}%
\centering
\subfigure{
\begin{minipage}[b]{0.475\textwidth}
\includegraphics[width=1\textwidth]{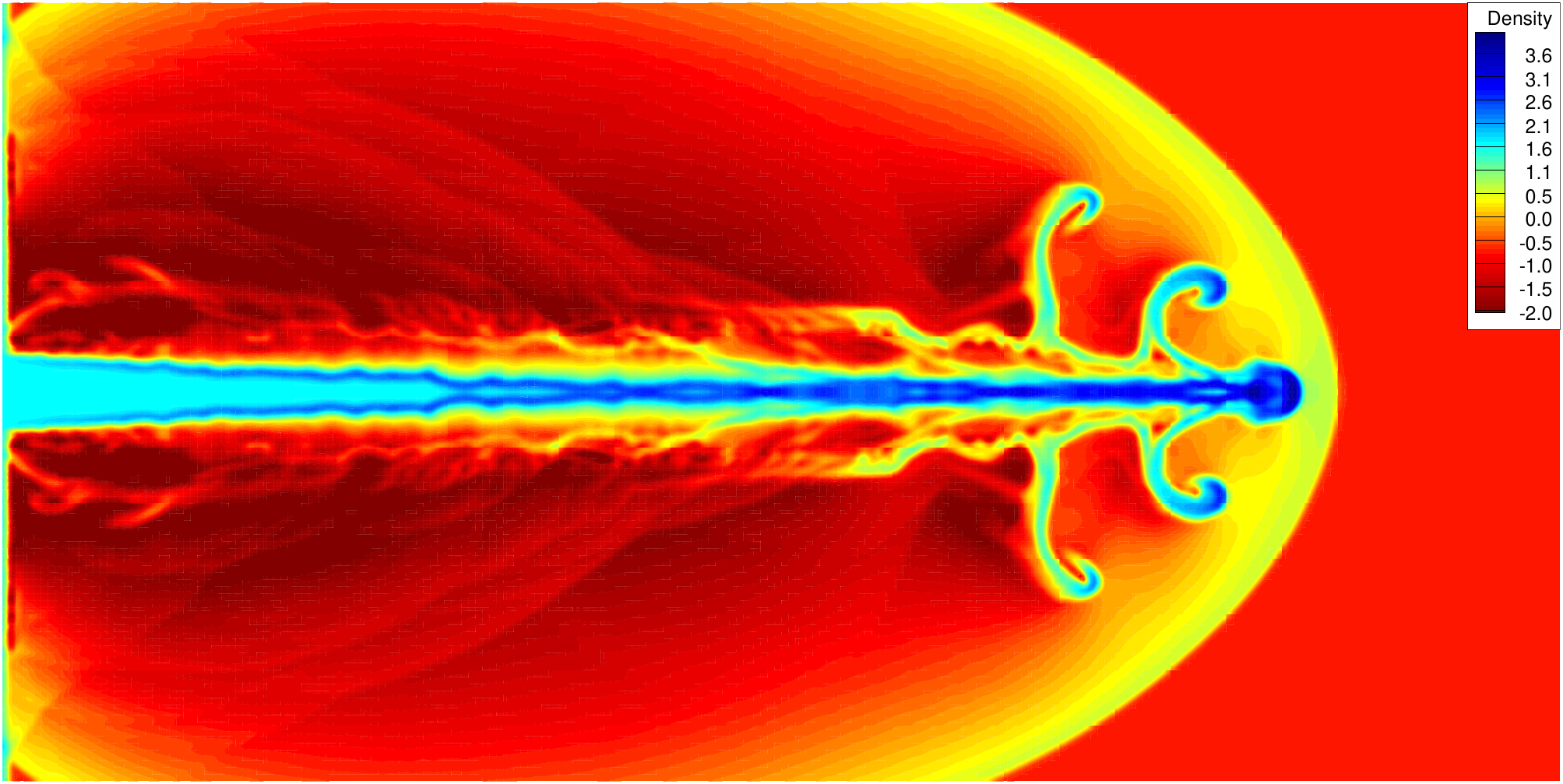} \\
\includegraphics[width=1\textwidth]{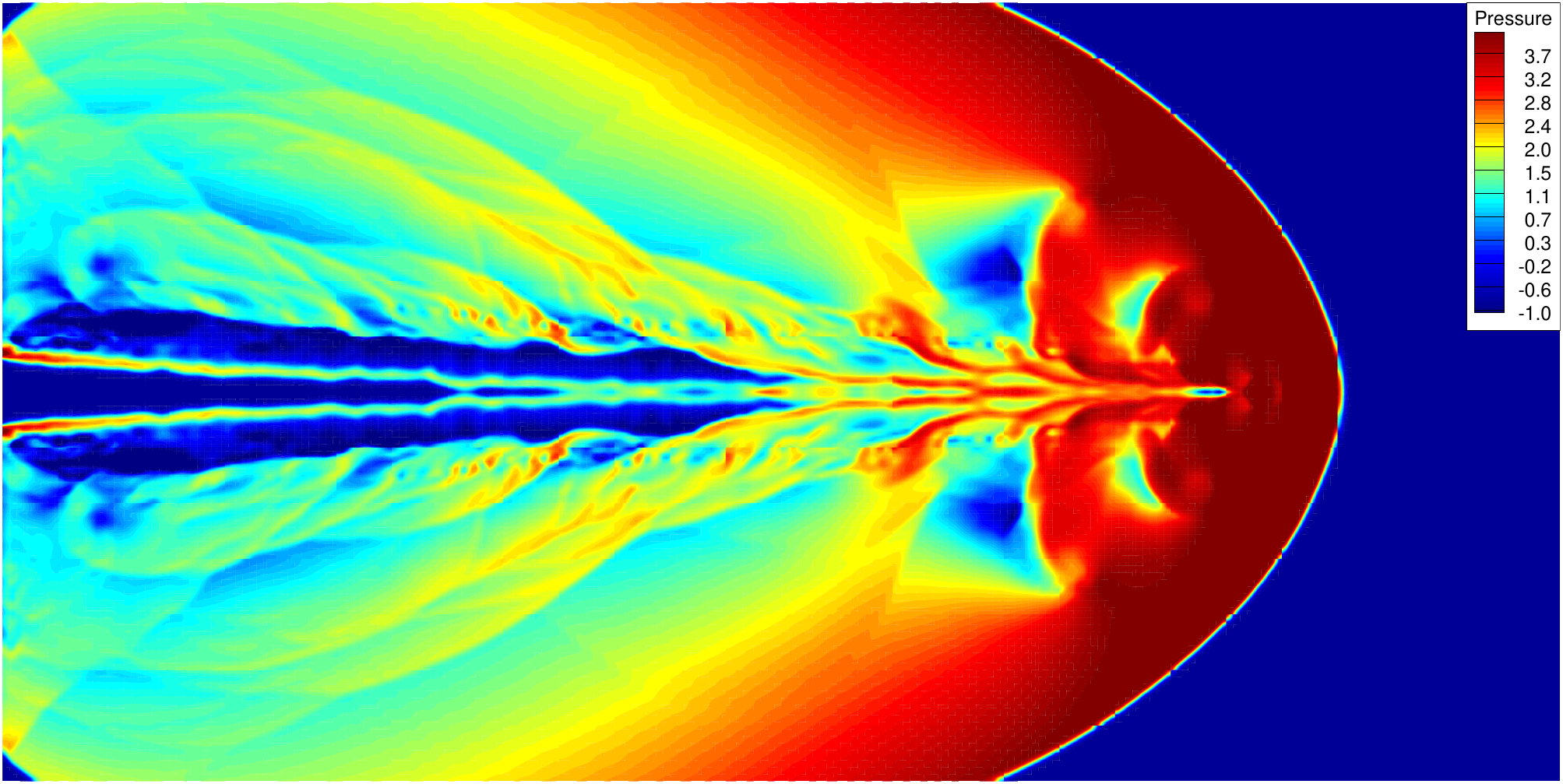}
\end{minipage}
}
\subfigure{
\begin{minipage}[b]{0.475\textwidth}
\includegraphics[width=1\textwidth]{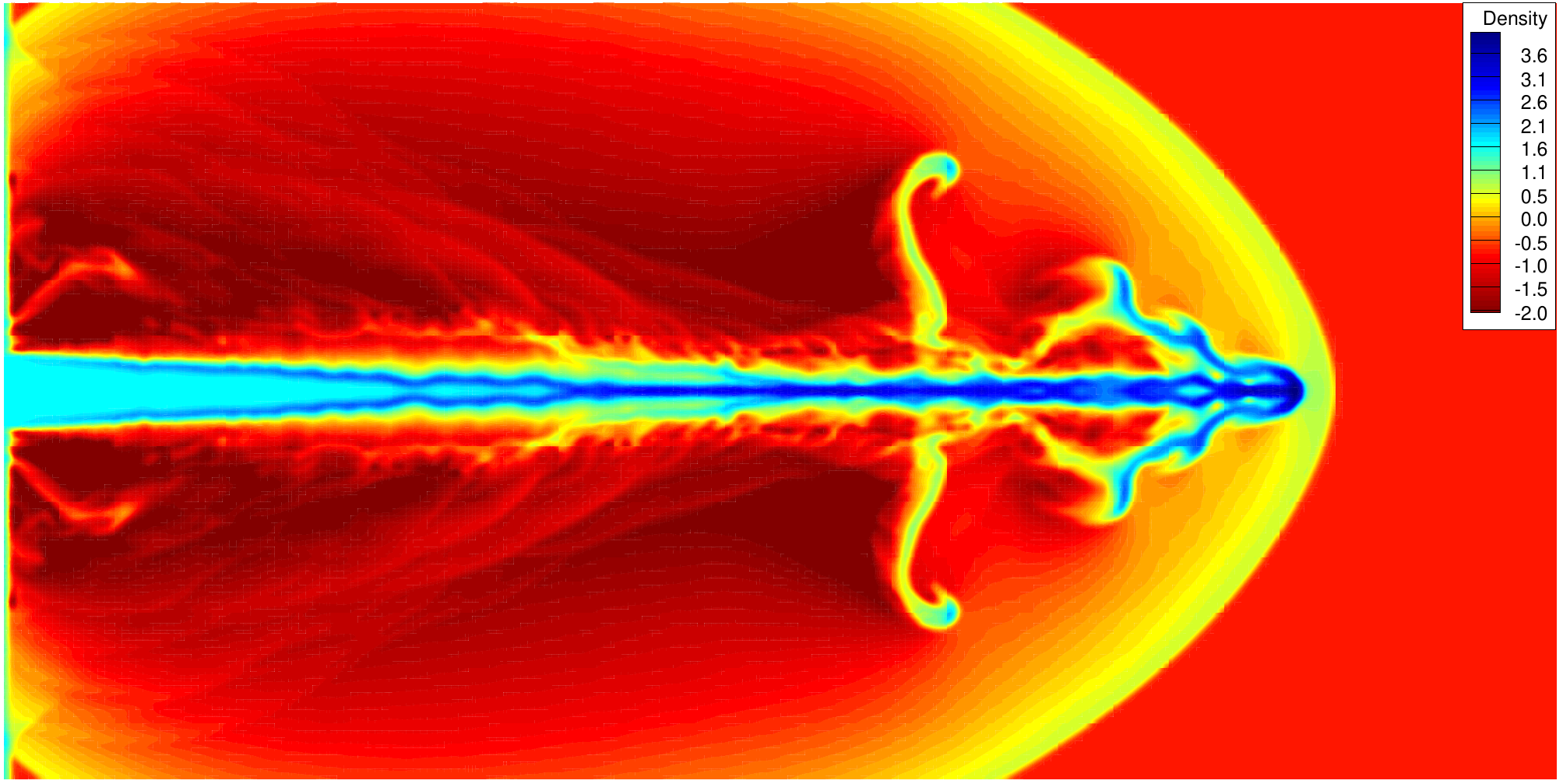} \\
\includegraphics[width=1\textwidth]{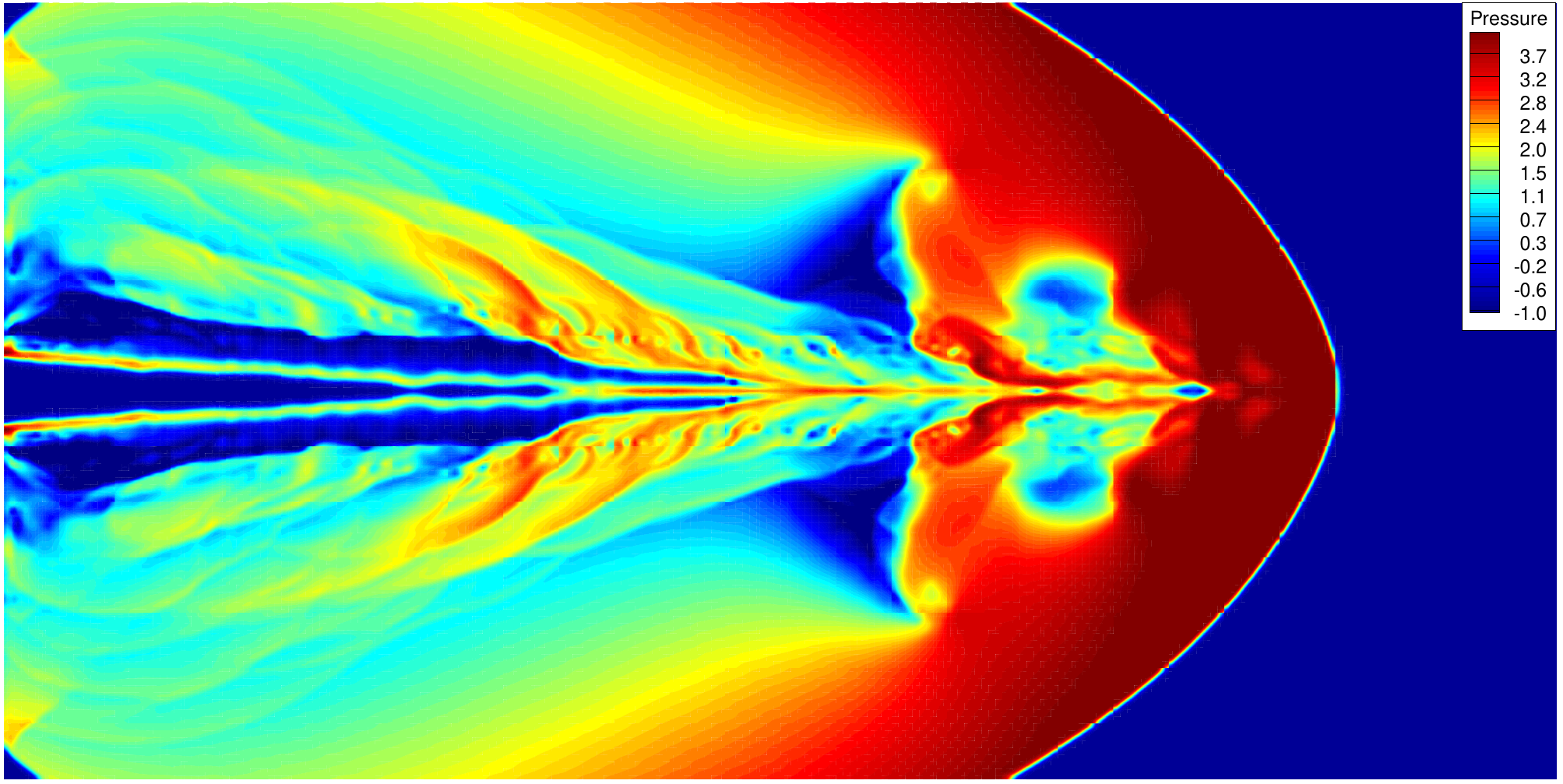}
\end{minipage}
}
  \caption{High Mach number astrophysical jet: density (top) and pressure (bottom) contours from TENO8-AA (left) and TENO10-AA (right) in logarithmic scale. The Mach number is 80 and the resolution is $448 \times 224$.}
 \label{Fig:80_Mach_number_astrophysical_jet}
\end{figure}

\begin{figure}%
\centering
\subfigure{
\begin{minipage}[b]{0.475\textwidth}
\includegraphics[width=1\textwidth]{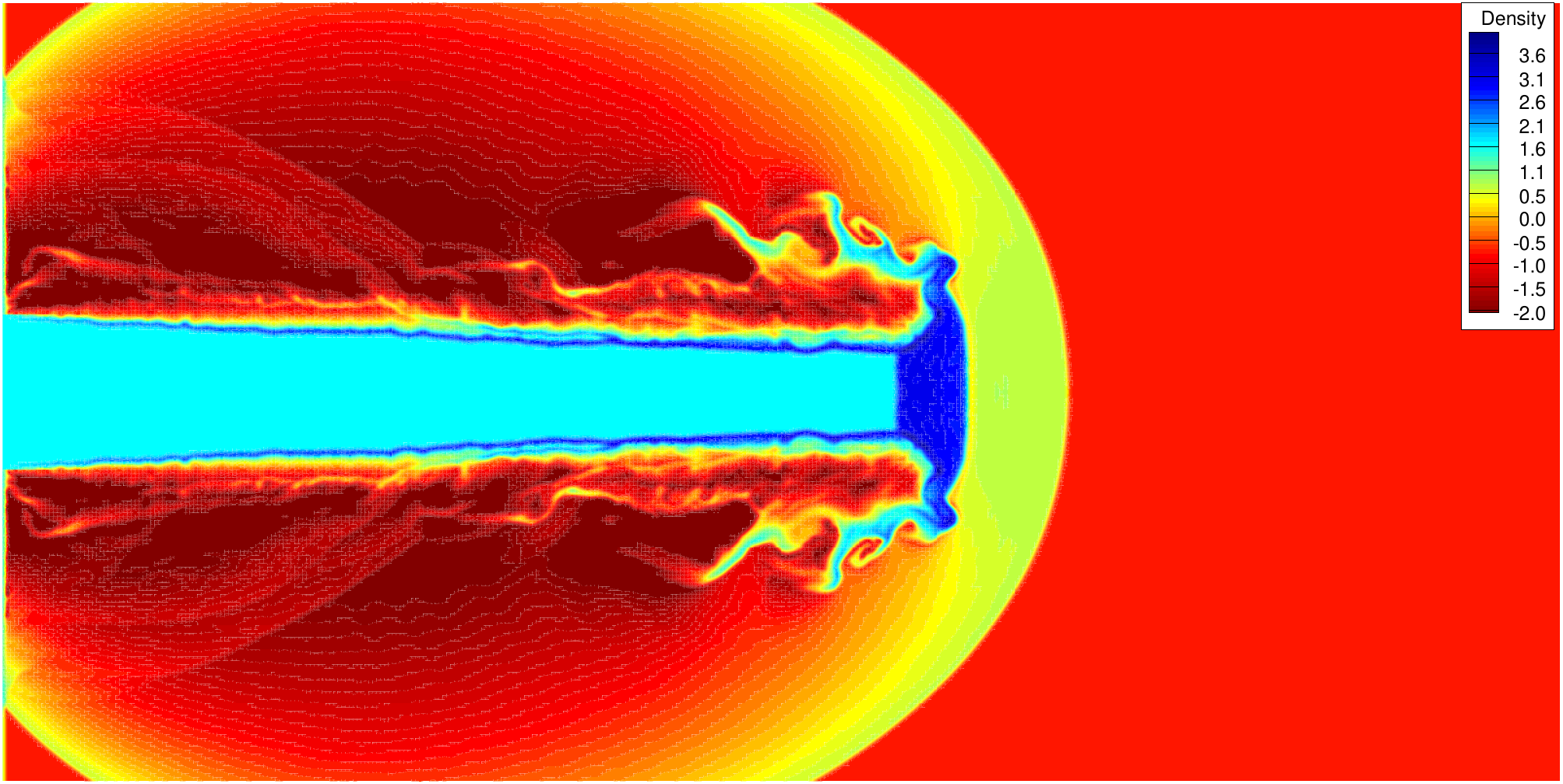} \\
\includegraphics[width=1\textwidth]{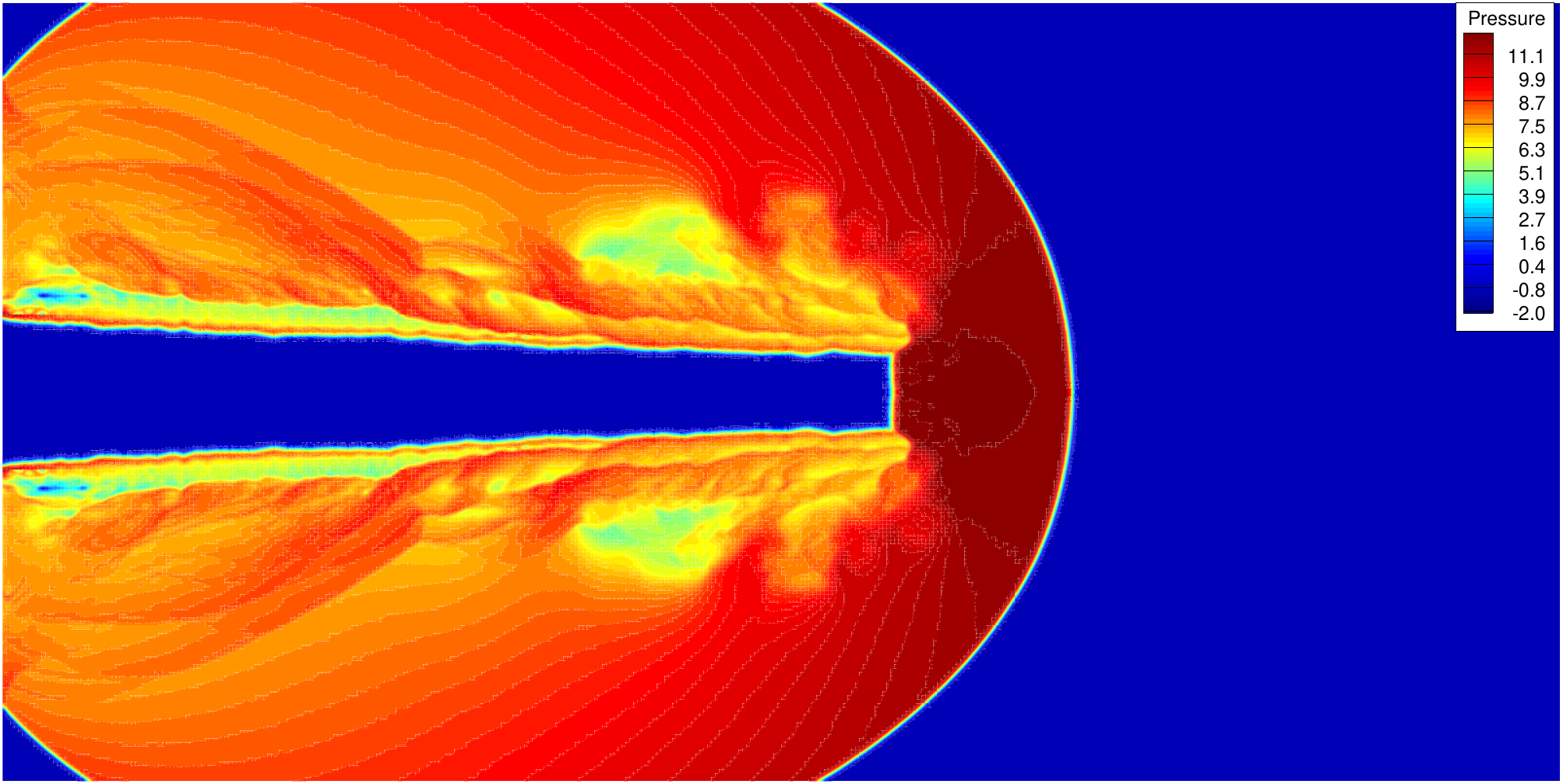}
\end{minipage}
}
\subfigure{
\begin{minipage}[b]{0.475\textwidth}
\includegraphics[width=1\textwidth]{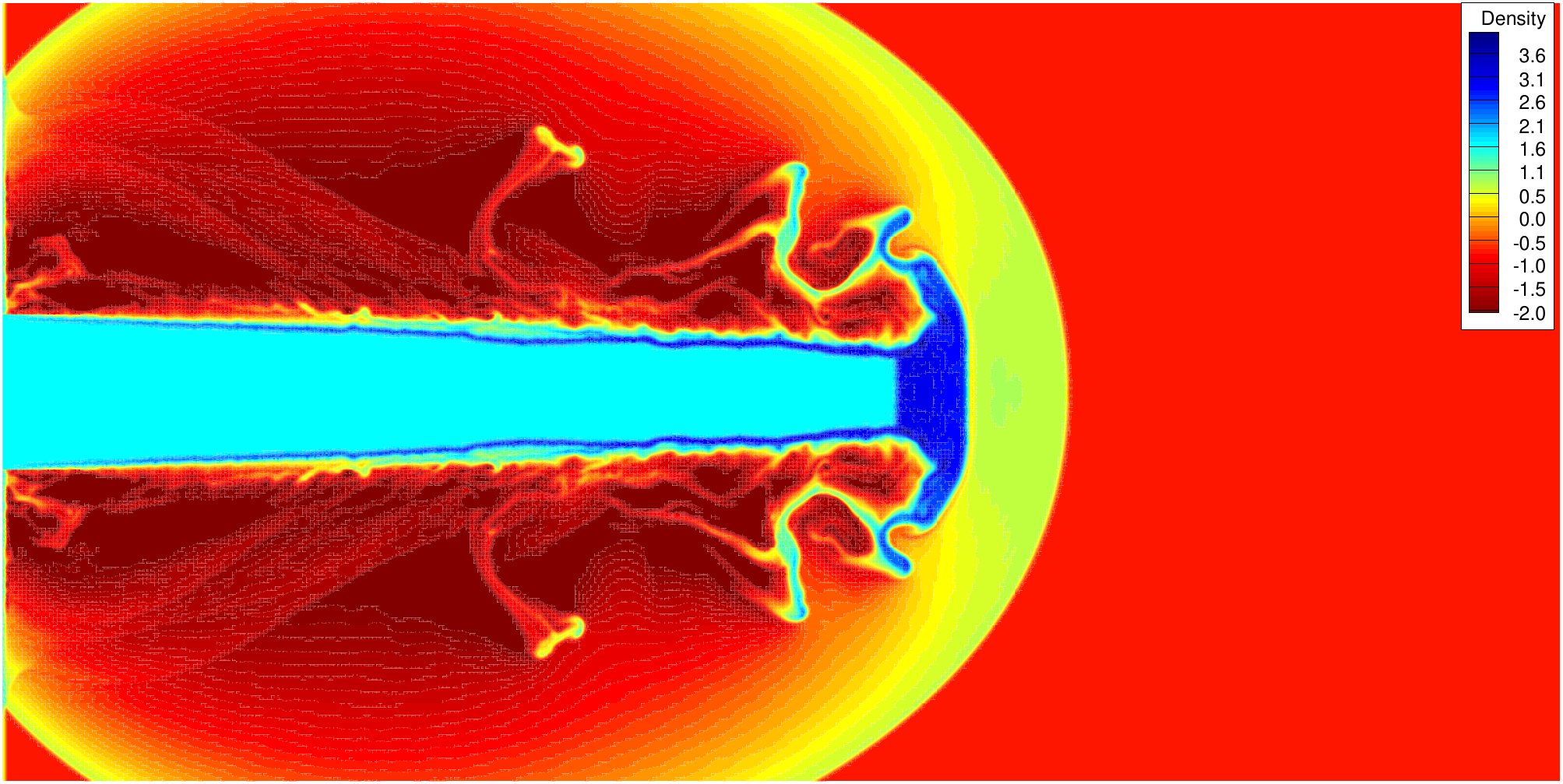} \\
\includegraphics[width=1\textwidth]{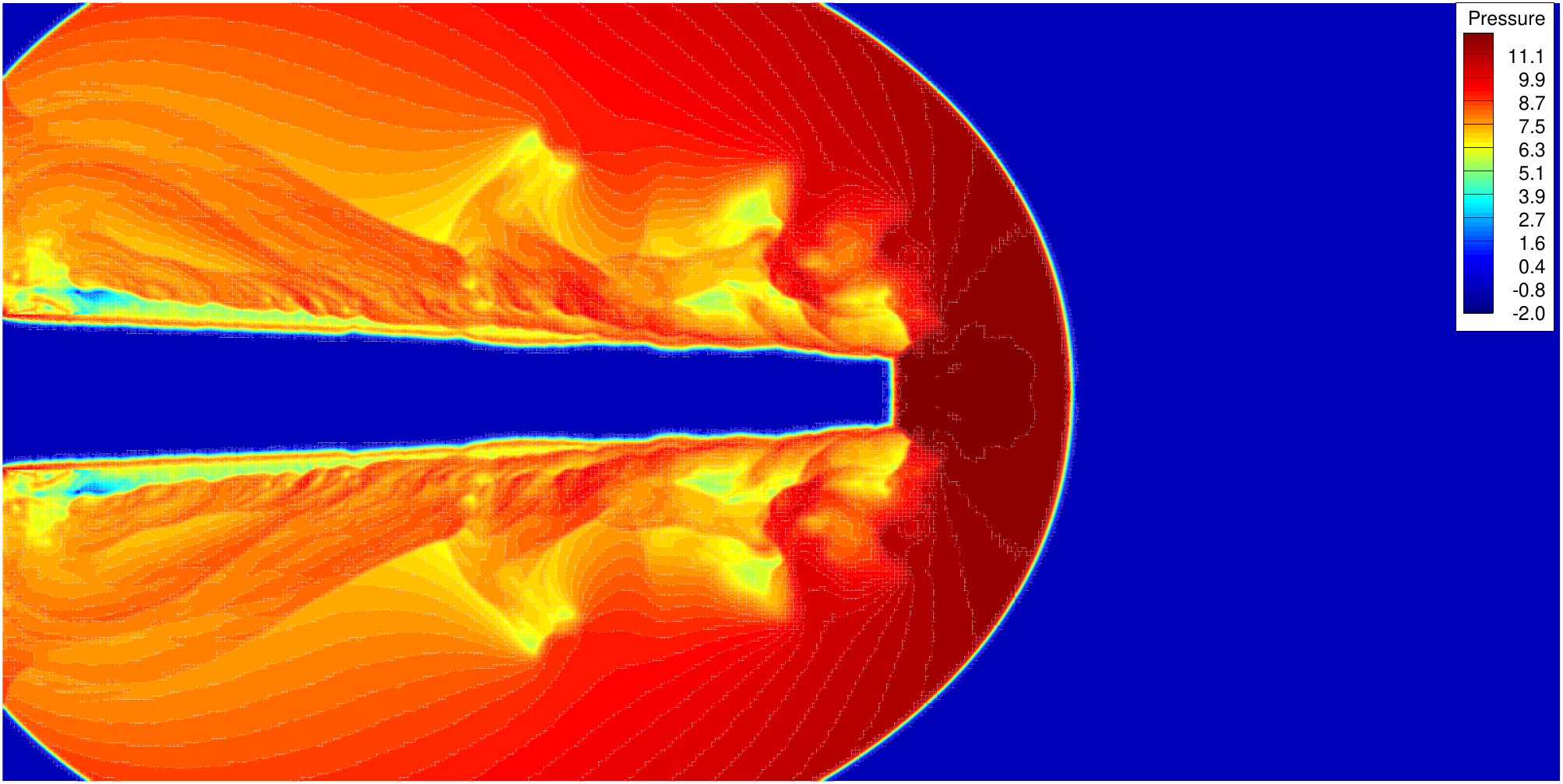}
\end{minipage}
}
  \caption{High Mach number astrophysical jet: density (top) and pressure (bottom) contours from TENO8-AA (left) and TENO10-AA (right) in logarithmic scale. The Mach number is 2000 and the resolution is $800 \times 400$.}
 \label{Fig:2000_Mach_number_astrophysical_jet}
\end{figure}

As shown in Fig.~\ref{Fig:80_Mach_number_astrophysical_jet} and Fig.~\ref{Fig:2000_Mach_number_astrophysical_jet}, while the resolved large-scale characteristic structures are similar, the proposed TENO8-AA and TENO10-AA schemes capture much richer small-scale structures than those from the fifth-order and seventh-order WENO scheme (see Fig.~5.4 of \cite{zhang2012positivity}, Fig.~7 of \cite{kotov2013comparative}), and that from the high-order discontinuous Galerkin scheme (see Fig.~4.5 and Fig.~4.6 of \cite{zhang2010positivity}). It is noted that the details of mushroom structures in the jet front are very sensitive to the numerical dissipation.
\subsection{3D Noh problem}
We consider the 3D Noh problem \cite{noh1987errors}, where the uniform implosion of an ideal gas leads to a shock of infinite strength and thus greatly challenges the numerical algorithms. The computational domain is $[0,0.256] \times [0,0.256] \times [0,0.256]$ and the ratio of the specific heats $\gamma = 5/3$. The initial condition is given as $(\rho, u, v, p) = (1, -x/r, -y/r, 10^{-6})$, where $r = \sqrt{x_i x_i}$ and the small number of $10^{-6}$ is adopted to prevent complex eigenvalues \cite{kawai2010assessment}. Symmetry conditions are imposed along $x_i = 0, \text{ } i=1,2,3$ while supersonic inflow conditions are imposed to the remaining boundaries with the velocity and pressure from the initial condition and the density from the analytical solution. For the spherical geometry, the shockwave moves radially outward with a constant speed of $1/3$. The analytical solution for the density in three dimensions is
 \begin{equation}
 \label{eq:density_noh}
\rho  = \left\{ {\begin{array}{*{20}{c}}
{64, }&{r < t/3,}\\
{{(1 + t/r)}^2, }&{r \ge t/3.}
\end{array}} \right.
 \end{equation}
In this case, the Rusanov scheme \cite{Jiang1996} is adopted for flux splitting. The mesh resolution is $64 \times 64 \times 64$ and the final simulation time is $t=0.6$.

\begin{figure}%
\centering
\includegraphics[width=0.375\textwidth]{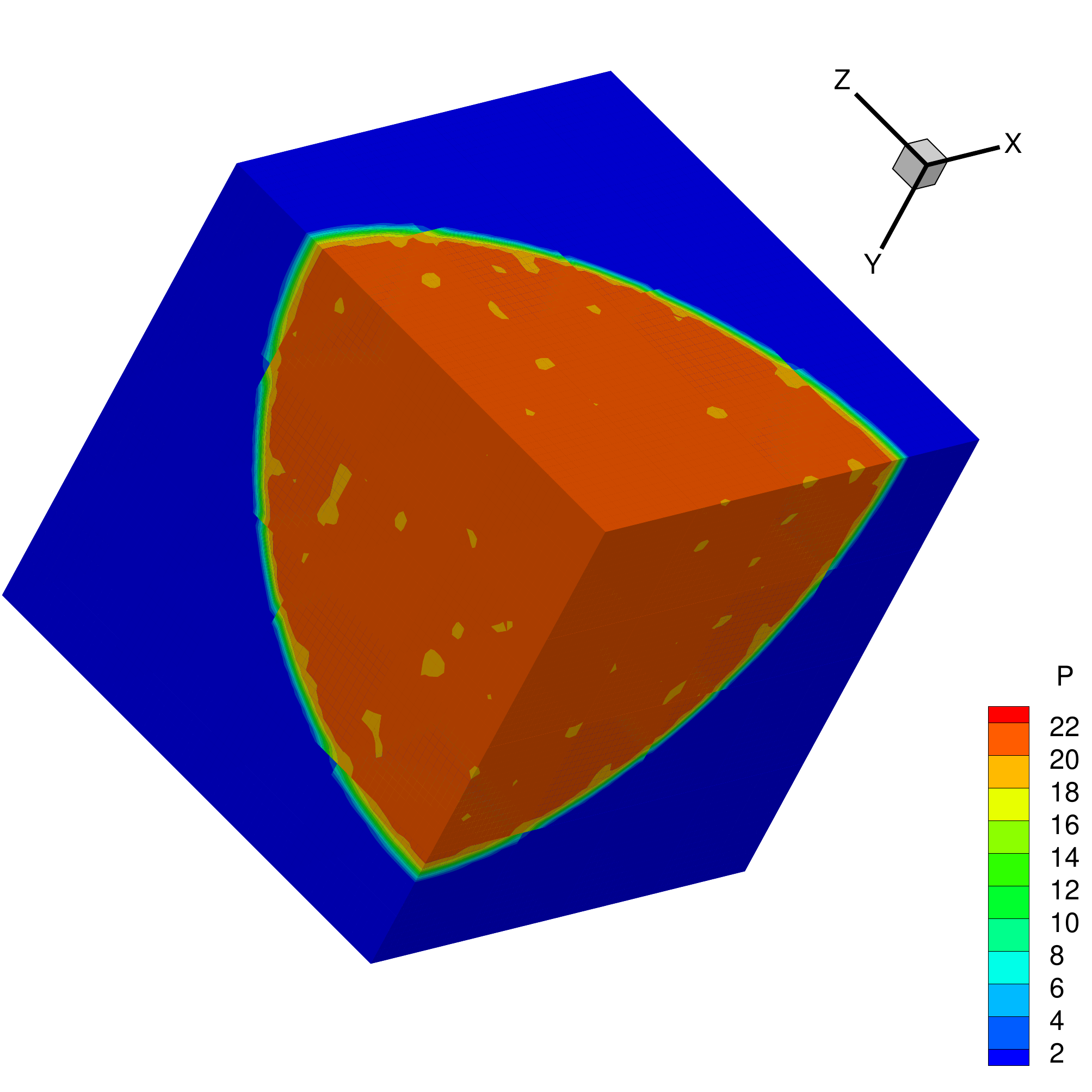}
\includegraphics[width=0.375\textwidth]{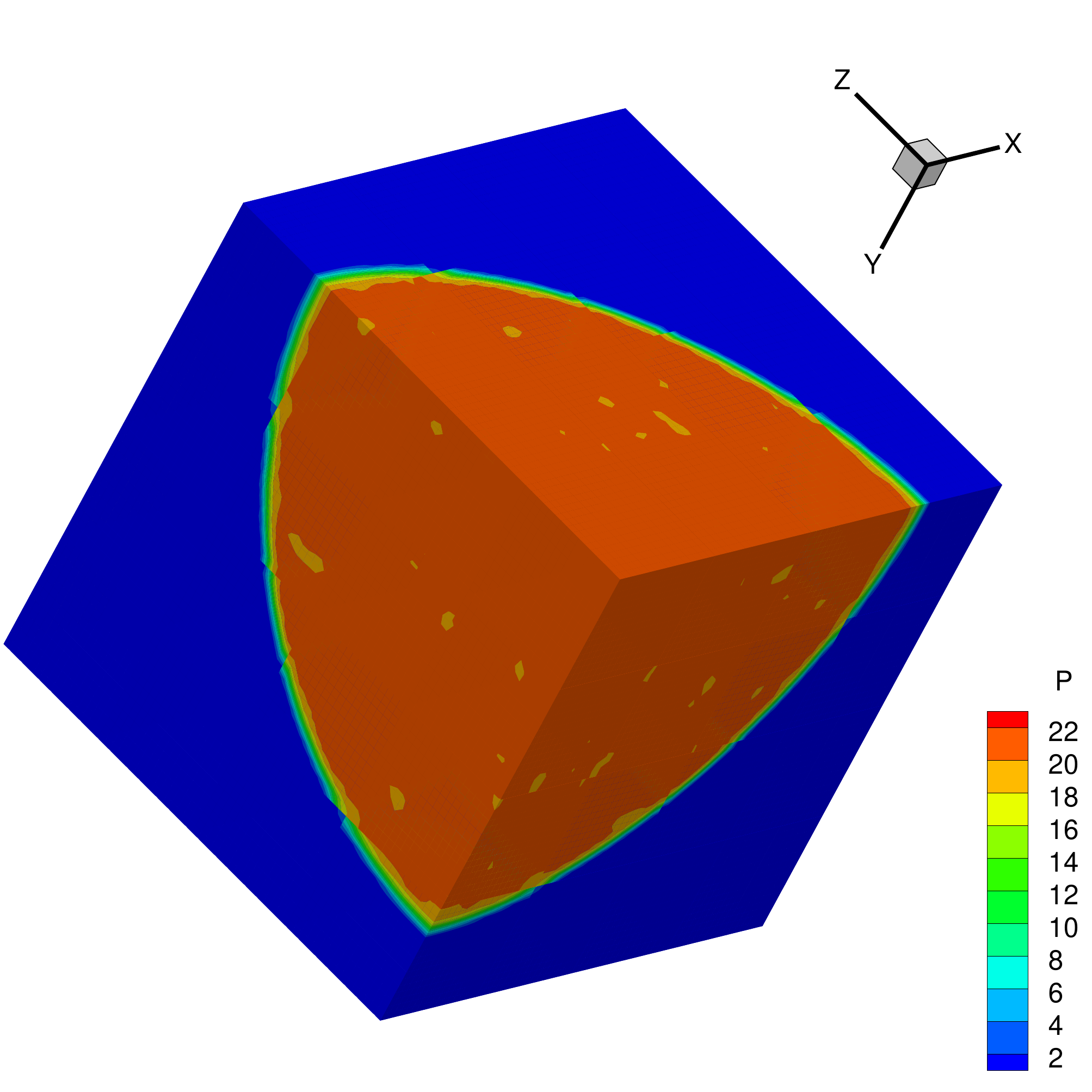} \\
\includegraphics[width=0.375\textwidth]{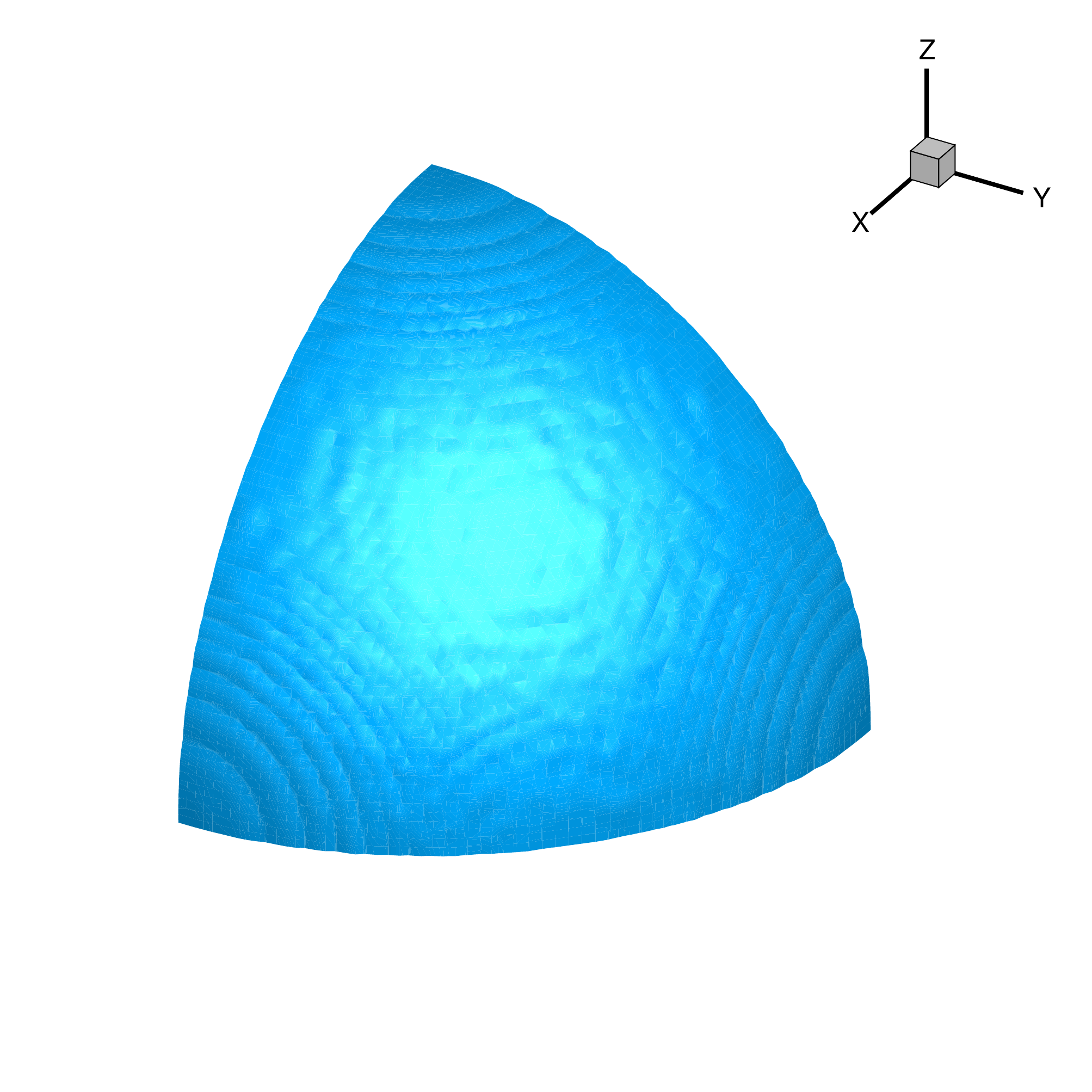}
\includegraphics[width=0.375\textwidth]{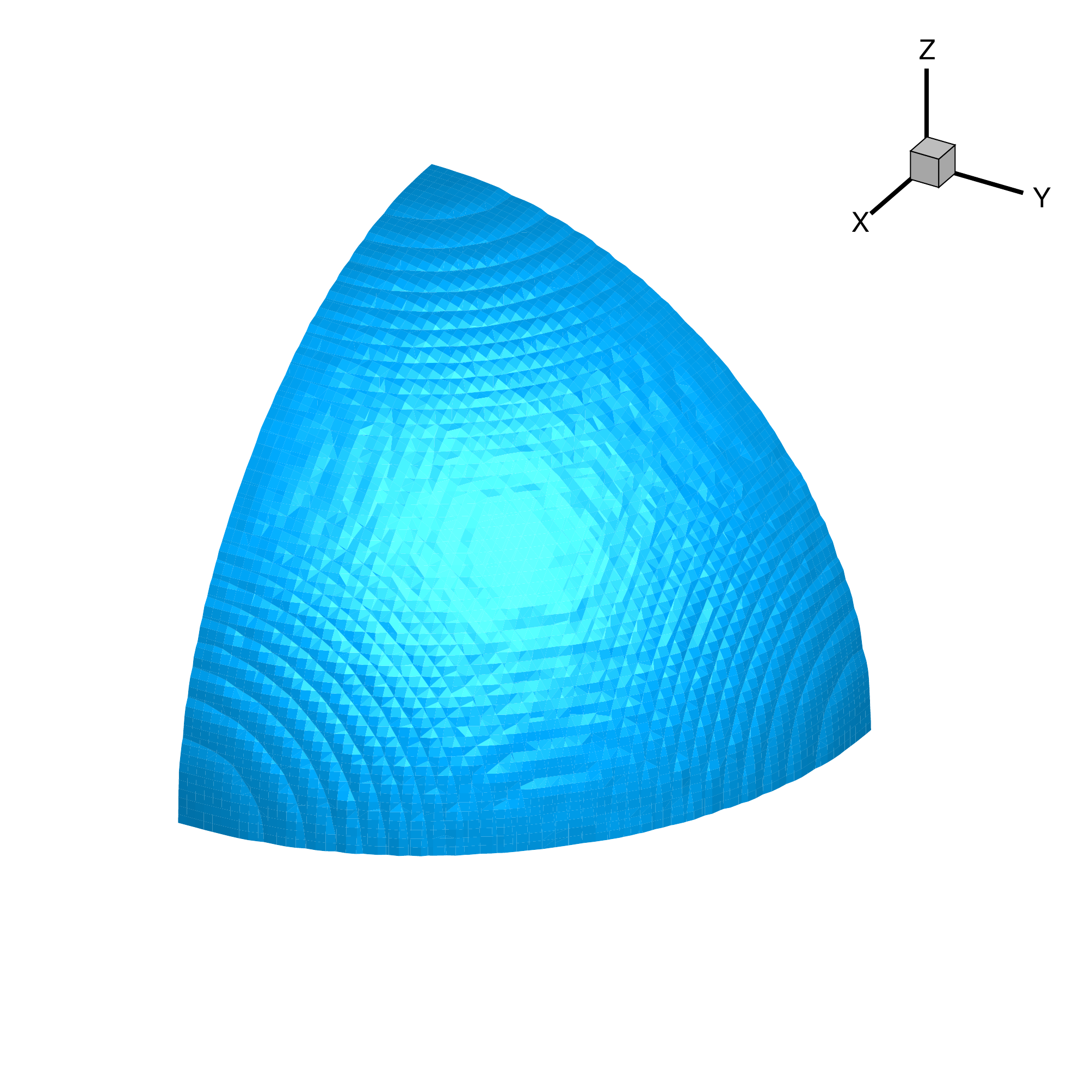} \\
\includegraphics[width=0.375\textwidth]{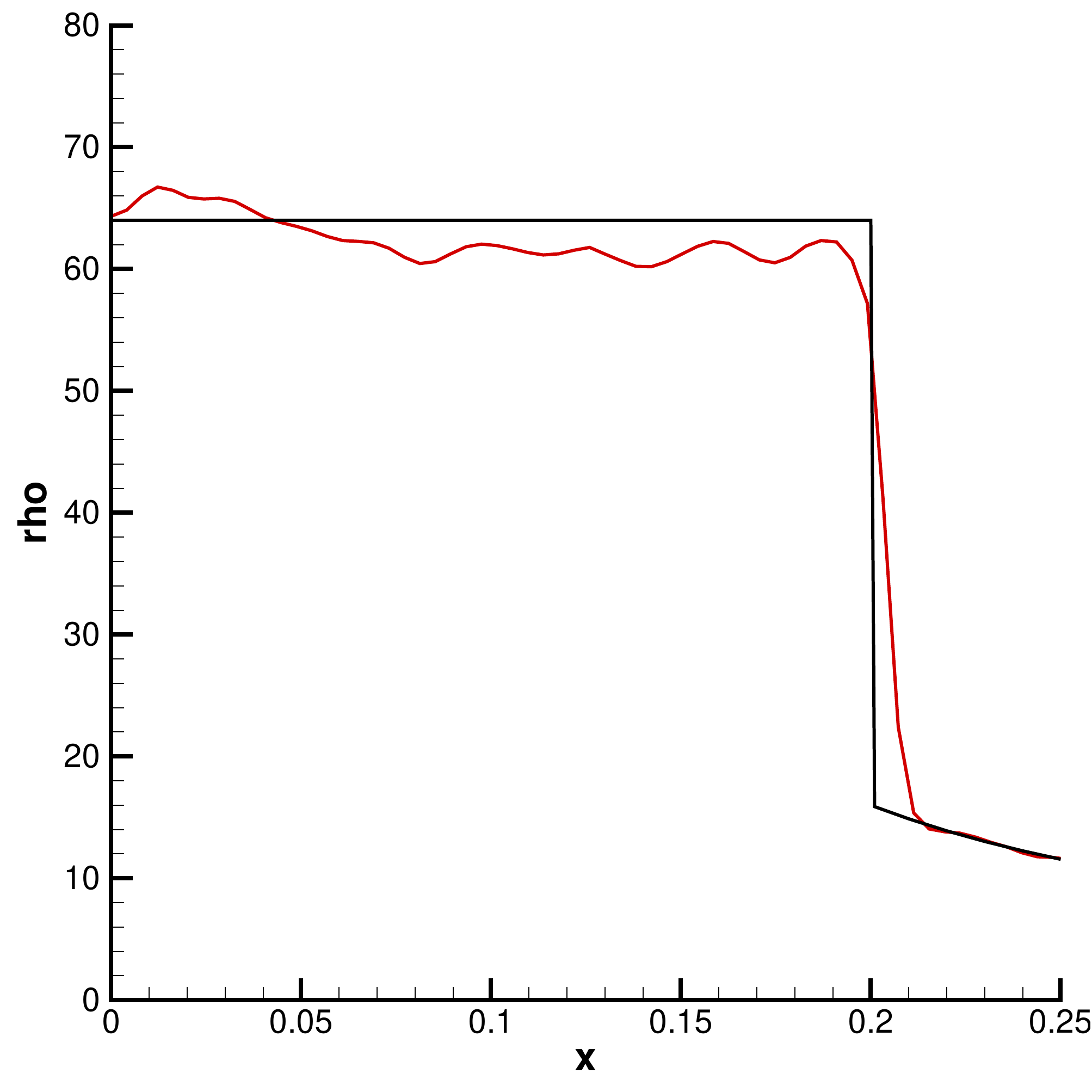}
\includegraphics[width=0.375\textwidth]{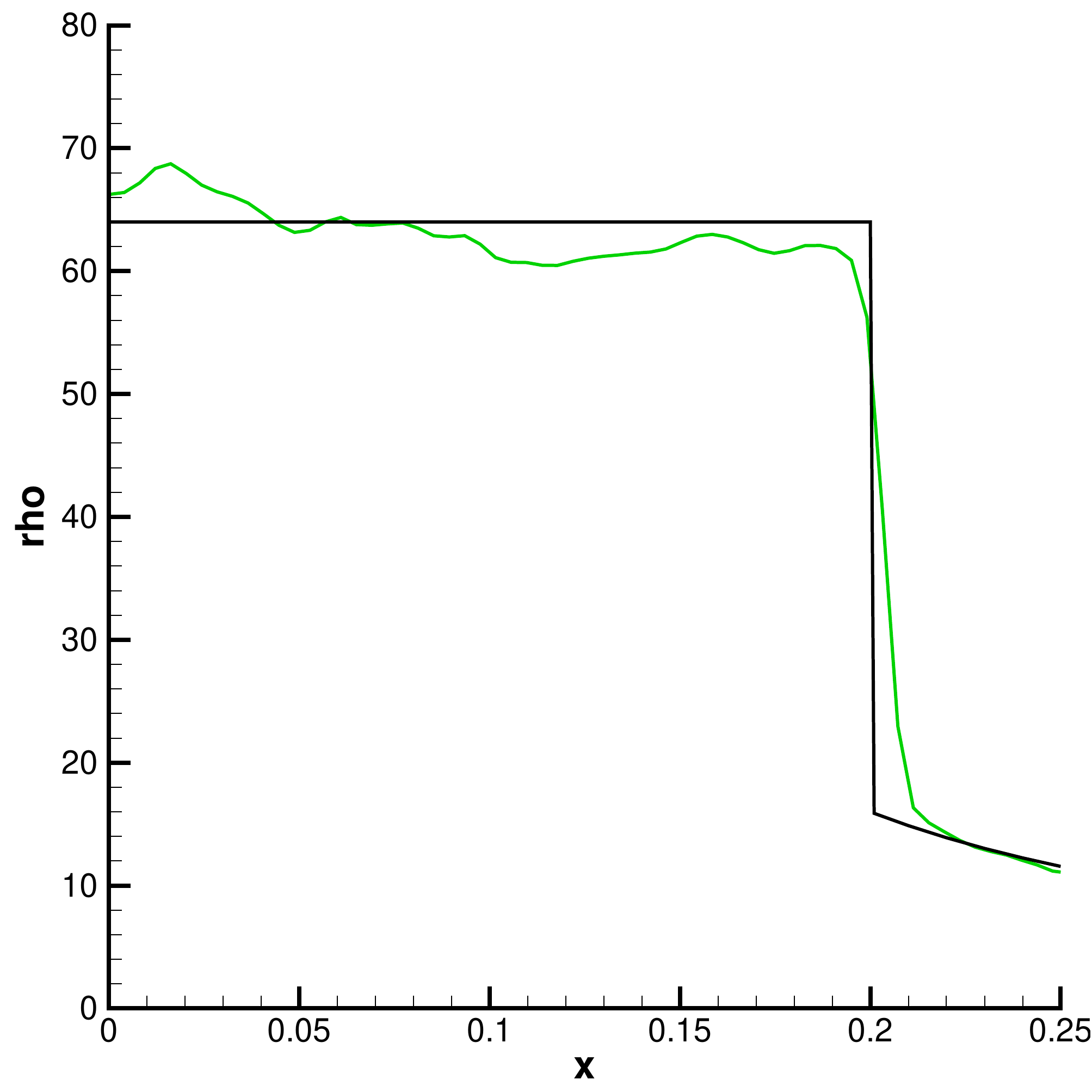} \\
  \caption{3D Noh problem: computed results from TENO8-AA (left) and TENO10-AA (right) at a resolution of $64 \times 64 \times 64$. From top to bottom: the pressure distribution, the iso-surface of density contour $\rho=20$, and the comparison between the computed density profile and the analytical solution.}
 \label{Fig:3D_noh}
\end{figure}

Fig.~\ref{Fig:3D_noh} gives the numerical results from TENO8-AA and TENO10-AA. From the pressure distribution and the iso-surface of density contours, it can be observed that the spherical shock-front shape is preserved well by both schemes. Moreover, compared with the results from other established high-order methods at a higher resolution of $128 \times 128$ (see Fig.10 in \cite{kawai2010assessment}), the present results are free from the wall-heating problem. Compared with that from the low-dissipation WENO-CU6-M1 scheme at the same resolution (see Fig.8 in \cite{hu2013positivity}), both TENO8-AA and TENO10-AA produce less density overshoots at the origin and the present results agree with the analytical solution much better. Also, TENO10-AA predicts more accurate post-shock density profile than TENO8-AA.
\section{Conclusions}

In this paper, a new family of very-high-order TENO schemes is developed. The major contributions of this paper are summarized as follows
\begin{itemize}

	\item A new framework to construct very-high-order TENO schemes is proposed. Based on the fact that the fifth-order TENO5 scheme is sufficiently robust for shock-capturing, the arbitrarily higher-order reconstruction is achieved by including more central-upwind large candidate stencils of incremental width in addition to the three small stencils of TENO5.

    \item A new ENO-like stencil selection strategy is developed. For smooth regions, the largest candidate stencil which features the best spectral property is assured to be deployed for the final reconstruction; for non-smooth regions, the newly proposed TENO scheme degenerates to lower-order reconstruction recursively. The nonlinear adaptation is invoked on the three small stencils when all the large candidates are unavailable for the final reconstruction.

    \item Based on the smoothness measurements of local flow scales with the first-order undivided difference, the cutoff parameter $C_T$ in the TENO weighting strategy is adapted in a way that minimum nonlinear dissipation is generated for high-wavenumber physical fluctuations while adequate dissipation is produced for shock-capturing. The set of built-in parameters are given explicitly.

    \item Without loss of generality, the eight- and ten-point TENO-AA schemes are constructed. A variety of benchmark cases are simulated to validate the performance of the proposed schemes. Numerical experiments demonstrate that, for conventional gas dynamics, the newly developed TENO-AA schemes feature low dissipation and are robust for discontinuity-capturing; for extreme simulations of very-high Mach number flows, the ENO-property is well preserved in combination with the positivity-preserving flux limiter.

\end{itemize}

Considering the good performance of the proposed schemes in benchmark simulations, they are promising for DNS and LES of practical engineering flows. {The in-house code with the present algorithms can be accessed upon reasonable requests to the corresponding author of this manuscript (the preliminary version can be found in \cite{hoppe2020modular})}.

\section*{Acknowledgements}

Lin Fu acknowledges the fund from Shenzhen Municipal Central Government Guides Local Science and Technology Development Special Funds Funded Projects (NO. 2021Szvup138).
\appendix
\renewcommand{\appendixname}{Appendix}
\section{Boundary condition implementation}
\label{sec:boundary_condition}

{
Considering a two-dimensional problem over the computational domain $ [0,X] \times [0,Y]$ and the time interval $[0,T]$, the space and time domain can be discretized into uniform intervals of size $\Delta x, \Delta y, \Delta t$. For a cell-centered pointwise variable $\phi$, $\phi_{i,j,k}=\phi |_{((i+\frac{1}{2})\Delta x,(j+\frac{1}{2})\Delta y,(k+\frac{1}{2})\Delta t)}$, where $0\le i \le (m-1), 0\le j \le (n-1), 0\le k\le (q-1)$. With regard to the physical variables $(\rho, u, v, p)$ in Euler problems, without loss of generality, we discuss different boundary condition implementations on the boundary $y=0$ (corresponding to the index $j=-1/2$) and $y=Y$ (corresponding to the index $j = n-1/2$) for any fixed $0\le i \le (m-1),0\le k\le (q-1)$. 

For the following interpretations, the left sides of the equations denote the ghost cell values to be defined, and the right sides denote the domain-interior values used to define the ghost cell values. $r$ denotes the ghost cell index, and the range depends on the need of the reconstruction schemes. For more complicated practical simulations with physical boundary conditions, e.g., the subsonic inflow, subsonic outflow, non-reflective boundaries, etc., the readers are referred to our open-source code as well as the corresponding journal paper for details, see the discussions of boundary condition implementation in section 4.3 of \cite{di2020htr}.

\subsection{Slip wall / symmetry boundary condition}

The $r$-th ghost cells on the sides of $y=0$ (first equation) and $y=Y$ (second equation) are defined as
\begin{equation}
\left\{
\begin{array}{rlc}
(\rho_{i,-r,k}, u_{i,-r,k}, v_{i,-r,k}, p_{i,-r,k})&=\quad (\rho_{i,r-1,k}, u_{i,r-1,k}, -v_{i,r-1,k}, p_{i,r-1,k}), \\ 
(\rho_{i,n-1+r,k}, u_{i,n-1+r,k}, v_{i,n-1+r,k}, p_{i,n-1+r,k})&=\quad (\rho_{i,n-r,k}, u_{i,n-r,k}, -v_{i,n-r,k}, p_{i,n-r,k}), \\
\end{array} \right. 
\end{equation}
where $r=1,2,3,\dots$, and the range depends on the need of the reconstruction schemes.

\subsection{No-slip wall boundary condition}

The $r$-th ghost cells on the sides of $y=0$ (first equation) and $y=Y$ (second equation) are defined as
\begin{equation}
\left\{
\begin{array}{rlc}
(\rho_{i,-r,k}, u_{i,-r,k}, v_{i,-r,k}, p_{i,-r,k})&=\quad (\rho_{i,r-1,k}, -u_{i,r-1,k}, -v_{i,r-1,k}, p_{i,r-1,k}), \\ 
(\rho_{i,n-1+r,k}, u_{i,n-1+r,k}, v_{i,n-1+r,k}, p_{i,n-1+r,k})&=\quad (\rho_{i,n-r,k}, -u_{i,n-r,k}, -v_{i,n-r,k}, p_{i,n-r,k}), \\
\end{array} \right.
\end{equation}
where $r=1,2,3,\dots$, and the range depends on the need of the reconstruction schemes.

\subsection{Dirichlet boundary condition}

Considering the Dirichlet boundary conditions given at $y=0$ (first equation) and $y=Y$ (second equation) as

$$\ \ \ \ \ \ (\rho_{i,-\frac{1}{2},k}, u_{i,-\frac{1}{2},k}, v_{i,-\frac{1}{2},k}, p_{i,-\frac{1}{2},k})=(\rho_{i,k} ^{\prime},u_{i,k}^{\prime},v_{i,k}^{\prime},p_{i,k}^{\prime}), $$
$$(\rho_{i,n-\frac{1}{2},k}, u_{i,n-\frac{1}{2},k}, v_{i,n-\frac{1}{2},k}, p_{i,n-\frac{1}{2},k})=(\rho_{i,k} ^{\prime \prime},u_{i,k}^{\prime \prime},v_{i,k}^{\prime \prime},p_{i,k}^{\prime \prime}),$$
the $r$-th ghost cells on the sides of $y=0$ (first equation) and $y=Y$ (second equation) are defined as
\begin{equation}
\left\{
\begin{array}{rlc}
(\rho_{i,-r,k}, u_{i,-r,k}, v_{i,-r,k}, p_{i,-r,k})&=\quad(2\rho ^{\prime}_{i,k}-\rho_{i,r-1,k},2u ^{\prime}_{i,k}-u_{i,r-1,k}, 2v ^{\prime}_{i,k}-v_{i,r-1,k}, 
\\& \ \ \ \ \ \ \ 2p ^{\prime}_{i,k}-p_{i,r-1,k}), \\
(\rho_{i,n-1+r,k}, u_{i,n-1+r,k}, v_{i,n-1+r,k}, p_{i,n-1+r,k})&=\quad(2\rho ^{\prime \prime}_{i,k}-\rho_{i,n-r,k}, 2u^{\prime \prime}_{i,k}-u_{i,n-r,k}, 2v^{\prime \prime}_{i,k}-v_{i,n-r,k},\\&\ \ \ \ \ \ \ 2p^{\prime \prime}_{i,k}- p_{i,n-r,k}),  \ 
\end{array} \right.
\end{equation}
where $r=1,2,3,\dots$, and the range depends on the need of the reconstruction schemes. Here, second-order approximation is enforced.

\subsection{Periodic boundary condition}

The $r$-th ghost cells on the sides of $y=0$ (first equation) and $y=Y$ (second equation) are defined as
\begin{equation}
\left\{
\begin{array}{rlc}
(\rho_{i,-r,k}, u_{i,-r,k}, v_{i,-r,k}, p_{i,-r,k})&=\quad(\rho_{i,n-r,k}, u_{i,n-r,k}, v_{i,n-r,k}, p_{i,n-r,k}),  \\
(\rho_{i,n-1+r,k}, u_{i,n-1+r,k}, v_{i,n-1+r,k}, p_{i,n-1+r,k})&=\quad(\rho_{i,r-1,k}, u_{i,r-1,k}, v_{i,r-1,k}, p_{i,r-1,k}), \\
\end{array} \right.
\end{equation}
where $r=1,2,3,\dots$, and the range depends on the need of the reconstruction schemes.
}



\bibliographystyle{acm}

\bibliography{explicit}

\end{document}